\documentclass[twoside,leqno]{article}

\usepackage[letterpaper]{geometry}

\usepackage{ltexpprt}

\usepackage{amsfonts,amsmath,amssymb}

\usepackage[colorlinks,citecolor=blue,linkcolor=blue,urlcolor=black,pagebackref]{hyperref}
\usepackage[T1]{fontenc}
\usepackage{mathdots}
\usepackage{bm,bbm}
\usepackage{url}
\usepackage{paralist}
\usepackage{enumerate}
\usepackage[normalem]{ulem}
\usepackage{xspace}
\xspaceaddexceptions{]\}}
\usepackage{comment}
\usepackage{mathtools}
\usepackage{tabu}
\usepackage{framed}
\usepackage{float,wrapfig}
\usepackage{subfigure}
\usepackage{tikz}
\usepackage[usenames,dvipsnames]{pstricks}

\usepackage[algo2e,linesnumbered,boxed,ruled,vlined]{algorithm2e}

\usepackage[capitalise]{cleveref}
\newtheorem{mytheorems}{Theorem}[section]
\newtheorem{thm}[mytheorems]{Theorem}
\newtheorem{lem}[mytheorems]{Lemma}
\newtheorem{prop}[mytheorems]{Proposition}
\newtheorem{cor}[mytheorems]{Corollary}

\newtheorem{definition}[mytheorems]{Definition}
\newtheorem{construction}[mytheorems]{Construction}
\newtheorem{remark}[mytheorems]{Remark}

\def\ShowAuthNotes{1}
\ifnum\ShowAuthNotes=1
\newcommand{\authnote}[2]{\ \\ \textcolor{red}{\parbox{0.9\linewidth}{[{\footnotesize {\bf #1:} { {#2}}}]}}\newline}
\else
\newcommand{\authnote}[2]{}
\fi

\renewcommand{\epsilon}{\varepsilon}
\newcommand{\eps}{\varepsilon}
\renewcommand{\Pr}{\operatorname*{\mathbf{Pr}}}
\newcommand{\Ex}{\operatorname*{\mathbf{E}}}

\newcommand{\poly}{\operatorname{\mathrm{poly}}}
\newcommand{\polylog}{\poly\log}
\newcommand{\per}{\operatorname{\mathsf{per}}}
\newcommand{\id}{\operatorname{\mathsf{id}}}
\newcommand{\lcp}{\operatorname{\mathsf{lcp}}}
\newcommand{\pos}{\operatorname{\mathsf{pos}}}
\newcommand{\suc}{\operatorname{\mathsf{succ}}}
\newcommand{\F}{\mathbbm{F}}

\newcommand{\N}{\mathbbm{N}}
\newcommand{\Q}{\mathbbm{Q}}
\newcommand{\sfS}{\mathsf{S}}
\newcommand{\sfU}{\mathsf{U}}
\newcommand{\caP}{\mathcal{P}}
\newcommand{\sfC}{\mathsf{C}}
\newcommand{\caC}{\mathcal{C}}

\def\dd{\mathinner{.\,.}}

\DeclarePairedDelimiter{\ket}{\lvert}{\rangle}

\DeclarePairedDelimiter{\set}{\{}{\}}

\DeclareMathOperator*{\argmin}{argmin}

\begin{document}

%
%\renewcommand\relatedversion{\thanks{The full version of the paper can be accessed at \protect\url{https://arxiv.org/abs/1902.09310}}} % Replace URL with link to full paper or comment out this line

%\setcounter{chapter}{2} % If you are doing your chapter as chapter one,
%\setcounter{section}{3} % comment these two lines out.

%\title{\Large SIAM/ACM Preprint Series Macros for Use With LaTeX\relatedversion}
\title{\Large Near-Optimal Quantum Algorithms for String Problems}

\author{
Shyan Akmal\thanks{{\url{naysh@mit.edu}}. Supported by NSF CCF-1909429 and a Siebel Scholarship.}\\MIT \and Ce Jin\thanks{\url{cejin@mit.edu}. Supported by an Akamai Presidential Fellowship and NSF CCF-2129139.}\\MIT
}

\date{}
\maketitle

	\begin{abstract}\small%\baselineskip=9pt
	We study quantum algorithms for several fundamental string problems, including \emph{Longest Common Substring}, \emph{Lexicographically Minimal String Rotation}, and \emph{Longest Square Substring}.
	These problems have been widely studied in the stringology literature since the 1970s, and are known to be solvable by near-linear time classical algorithms. 
	In this work, we give quantum algorithms for these problems with \emph{near-optimal} query complexities and time complexities.  Specifically, we show that:
	\begin{itemize}
	    \item \emph{Longest Common Substring} can be solved by a quantum algorithm in $\tilde O(n^{2/3})$ time, improving upon the recent $\tilde O(n^{5/6})$-time algorithm by Le Gall and Seddighin (2020). Our algorithm uses the MNRS quantum walk framework, together with a careful combination of string synchronizing sets (Kempa and Kociumaka, 2019) and generalized difference covers. 
	    \item  \emph{Lexicographically Minimal String Rotation} can be solved by a quantum algorithm in $n^{1/2 + o(1)}$ time, improving upon the recent $\tilde O(n^{3/4})$-time algorithm by Wang and Ying (2020).
	    We design our algorithm by first giving a new classical divide-and-conquer algorithm in near-linear time based on exclusion rules, and then  speeding it up quadratically using nested Grover search and quantum minimum finding. 
	    \item  \emph{Longest Square Substring} can be solved by a quantum algorithm in $\tilde O(\sqrt{n})$ time. Our algorithm is an adaptation of the algorithm by Le Gall and Seddighin (2020) for the Longest Palindromic Substring problem, but uses additional techniques to overcome the difficulty that binary search no longer applies.
	\end{itemize}
	Our techniques naturally extend to other related string problems, such as Longest Repeated Substring, Longest Lyndon Substring, and Minimal Suffix.
	\end{abstract}

\section{Introduction}

The study of string processing algorithms is an important area of research in theoretical computer science, with applications in numerous fields including bioinformatics, data mining, plagiarism detection, etc.
Many fundamental problems in this area have been known to have linear-time algorithms since over 40 years ago. Examples include \emph{Exact String Matching} \cite{DBLP:journals/siamcomp/KnuthMP77,DBLP:journals/ibmrd/KarpR87}, \emph{Longest Common Substring} \cite{DBLP:conf/focs/Weiner73,DBLP:conf/focs/Farach97,DBLP:conf/csr/BabenkoS08}, and \emph{(Lexicographically) Minimal String Rotation} \cite{DBLP:journals/ipl/Booth80,DBLP:journals/jal/Shiloach81,DBLP:journals/jal/Duval83}. 
These problems have also been studied extensively in the context of data structures, parallel algorithms, and low-space algorithms. 

More recently, there has been growing interest in developing efficient \emph{quantum algorithms} for these basic string problems. Given quantum query access to the input strings (defined in \cref{sec:model}), it is sometimes possible to solve such problems in \emph{sublinear} query complexity and time complexity.
The earliest such result was given by Ramesh and Vinay \cite{matching}, who combined Vishkin's deterministic sampling technique \cite{DBLP:journals/siamcomp/Vishkin91} with Grover search \cite{DBLP:conf/stoc/Grover96} to obtain a quantum algorithm for the Exact String Matching problem with near-optimal $\tilde O(\sqrt{n})$ time complexity\footnote{Throughout this paper, $\tilde O(\cdot )$ hides $\polylog n$ factors where $n$ denotes the input length, and $\tilde \Omega(\cdot),\tilde \Theta(\cdot)$ are defined analogously. In particular, $\tilde O(1)$ means $O(\polylog n)$.}. 
More recently, Le Gall and Seddighin \cite{legall} obtained sublinear-time quantum algorithms for various string problems, among them an $\tilde O(n^{5/6})$-time algorithm for Longest Common Substring (LCS) and an $\tilde O(\sqrt{n})$-time algorithm for Longest Palindromic Substring (LPS).
In developing these algorithms, they applied the quantum Exact String Matching algorithm \cite{matching} 
and Ambainis' Element Distinctness algorithm \cite{DBLP:journals/siamcomp/Ambainis07} as subroutines, and used periodicity arguments to reduce the number of candidate solutions to be checked.
Another recent work by Wang and Ying \cite{ying} showed that
Minimal String Rotation can be solved in $\tilde O(n^{3/4})$ quantum time. Their algorithm was also based on quantum search primitives (including Grover search  and quantum minimum finding \cite{minimumfinding}) and techniques borrowed from parallel string algorithms \cite{DBLP:conf/paa/ApostolicoIP87,DBLP:journals/siamcomp/Vishkin91,DBLP:journals/tcs/IliopoulosS92}.

On the lower bound side, it has been shown that Longest Common Substring requires $\tilde \Omega(n^{2/3})$ quantum query complexity (by a reduction \cite{legall} from the Element Distinctness problem \cite{DBLP:journals/jacm/AaronsonS04,DBLP:journals/toc/Kutin05,DBLP:journals/toc/Ambainis05}), and that Exact String Matching, Minimal String Rotation, and Longest Palindromic Substring all require $\Omega(\sqrt{n})$ quantum query complexity (by reductions \cite{legall,ying} from the unstructured search problem \cite{DBLP:journals/siamcomp/BennettBBV97}). 
Le Gall and Seddighin \cite{legall} observed that although the classical algorithms for LCS and LPS are almost the same (both based on suffix trees \cite{DBLP:conf/focs/Weiner73}), the latter problem (with time complexity $\tilde \Theta(\sqrt{n})$) is strictly easier than the former (with an $\tilde \Omega(n^{2/3})$ lower bound) in the quantum query model.

Despite these results, our knowledge about the quantum computational complexities of basic string problems is far from complete. 
For the LCS problem and the Minimal String Rotation problem mentioned above, there are $n^{\Omega(1)}$ gaps between current upper bounds and lower bounds.  Better upper bounds are only known in special cases: Le Gall and Seddighin \cite{legall} gave an $\tilde O(n^{2/3})$-time algorithm for $(1-\eps)$-approximating LCS in non-repetitive strings, matching the query lower bound in this setting. Wang and Ying \cite{ying} gave an $\tilde O(\sqrt{n})$-time algorithm for Minimum String Rotation in randomly generated strings, and showed a matching average-case query lower bound. However, these algorithms do not immediately extend to the general cases. 
Moreover, there remain many other string problems which have near-linear time classical algorithms with no known quantum speed-up.

\subsection{Our Results}
In this work, we develop new quantum query algorithms for many fundamental string problems. \emph{All our algorithms are near-optimal and time-efficient}: they have time complexities that match the corresponding query complexity lower bounds up to $n^{o(1)}$ factors. 
In particular, we close the gaps for Longest Common Substring and Minimal String Rotation left open in previous work \cite{legall,ying}. We summarize our contributions (together with some earlier results) in \cref{fig:table}. See \Cref{sec:problems} for formal definitions of the studied problems.

\begin{figure*}
    \centering
\begin{tabular}{|c|c|c|c|c|}
\hline \hline
\rule{0pt}{10pt}
{\bf Problem} & {\bf Time UB}  & {\bf Reference }& {\bf Query LB} \\[0.8pt]
\hline \rule{0pt}{17pt} \hspace{-8pt}
Longest Common Substring & 
 $\begin{matrix}\tilde O(n^{5/6})   \\ \tilde O(n^{2/3})\end{matrix}$
& $\begin{matrix}\text{\cite{legall}} \\ \text{\textbf{This work} (\Cref{thm:lcs-main})}\end{matrix}$ & $\tilde \Omega(n^{2/3})$  \\
Longest Repeated Substring & 
$\tilde O(n^{2/3})$
& \textbf{This work} (\Cref{remark:repeat}) & $\tilde \Omega(n^{2/3})$  \\[1pt]
\hline
 \rule{0pt}{17pt}Minimal String Rotation&
$\begin{matrix}\tilde O(n^{3/4}) \\ n^{1/2+o(1)} \end{matrix}$
 & $\begin{matrix}\text{\cite{ying}} \\ \text{\textbf{This work} (\Cref{thm:lexico-main})}\end{matrix}$ & $ \Omega(\sqrt{n})$  \\
Minimal Suffix& $n^{1/2+o(1)}$ & \textbf{This work} (\Cref{thm:lexico-main})&  $ \Omega(\sqrt{n})$  \\
Maximal Suffix& $n^{1/2+o(1)}$ & \textbf{This work} (\Cref{thm:lexico-main})& $ \Omega(\sqrt{n})$   \\
Longest Lyndon Substring & $n^{1/2+o(1)}$ & \textbf{This work} (\Cref{thm:long-lyndon})&  $ \Omega(\sqrt{n})$  \\[1pt]
\hline \rule{0pt}{11pt}
Longest Palindromic Substring & $\tilde{O}(\sqrt{n})$ & \cite{legall} & $ \Omega(\sqrt{n})$  \\
Longest Square Substring& $\tilde{O}(\sqrt{n})$ & \textbf{This work} (\cref{thm:lss-main}) & $ \Omega(\sqrt{n})$  \\[1pt]
\hline 
\hline
\end{tabular}
    \caption{Near-optimal quantum algorithms for string problems (see definitions in \cref{sec:problems}). Problems are grouped based on similarity. All problems listed here have near-linear time classical algorithms.}
    \label{fig:table}
\end{figure*}

\subsection{Technical Overview}
We give high-level overviews of our quantum algorithms for Longest Common Substring (LCS), Minimal String Rotation, and Longest Square Substring.
\subsubsection{Longest Common Substring}
\label{sec:overview-lcs}
We consider the decision version of LCS with threshold length $d$: given two length-$n$ input strings $s,t$, decide whether they have a common substring of length $d$.

Le Gall and Seddighin \cite[Section 3.1.1]{legall} observed a simple reduction from this decision problem to the (bipartite version of) Element Distinctness problem,  which asks whether the two input lists $A,B$ contain a pair of identical items $A_i=B_j$. Ambainis \cite{DBLP:journals/siamcomp/Ambainis07} gave a comparison-based algorithm for this problem in $\tilde O(n^{2/3}\cdot T)$ time, where $T$ denotes the time complexity of comparing two items. 
In the LCS problem of threshold length $d$, each item is a length-$d$ substring  of $s$ or $t$ (specified by the starting position), and the lexicographical order between two length-$d$ substrings can be compared in $T=\tilde O(\sqrt{d})$ using binary search and Grover search (see \cref{lem:lcp-grover}). Hence, this problem can be solved in $\tilde O(n^{2/3}\cdot \sqrt{d})$ time.

\paragraph*{The anchoring technique.} The inefficiency of the algorithm described above comes from the fact that there are $n-d+1 = \Omega(n)$ positions to be considered in each input string. This seems rather unnecessary for larger $d$, since intuitively there is a lot of redundancy from the large overlap between these length-$d$ substrings. 
This is the idea behind the so-called \emph{anchoring} technique, which has been widely applied in designing classical algorithms for various versions of the LCS problem \cite{DBLP:conf/cpm/StarikovskayaV13,DBLP:conf/cpm/Charalampopoulos18,DBLP:conf/esa/AmirCPR19,DBLP:journals/algorithmica/AmirCPR20,DBLP:conf/cpm/Nun0KK20,icalp20,lcs2021}.
In this technique, we carefully pick subsets $C_1,C_2 \subseteq [n]$ of \emph{anchors}, such that in a YES input instance there must exist an \emph{anchored common substring}, i.e., a common string with occurrences $s[i\dd i+d)=t[j\dd j+d)$ and a shift $0\le h <d$ such that $i+h\in C_1$ and $j+h \in C_2$. 
Then, the task reduces to the \emph{Two String Families LCP problem} \cite{DBLP:conf/cpm/Charalampopoulos18}, where we want to find a pair of anchors $i'\in C_1, j'\in C_2$ that can be extended in both directions to get a length-$d$ common substring, or equivalently,  the longest common prefix of $s[i'\dd],t[j'\dd]$ and the longest common suffix of $s[\dd i'-1], t[\dd j'-1]$ have total length at least $d$.  Intuitively, finding a smaller set of anchors would make our algorithm have better running time. 

\paragraph*{Small and explicit anchor sets.}
One can construct such anchor sets based on \emph{difference covers} 
\cite{DBLP:conf/cpm/BurkhardtK03,DBLP:journals/tocs/Maekawa85}, with size $|C_1|,|C_2|\le n/\sqrt{d}$. The construction is very simple and explicit (see \cref{sec:diffcover}), and is oblivious to the content of the input strings (in fact, it just consists of several arithmetic progressions of fixed lengths).
In comparison, there exist much smaller constructions if the anchors are allowed to depend on the input strings: for example, in their time-space tradeoff algorithm for LCS, Ben-Nun, Golan, Kociumaka, and Kraus \cite{DBLP:conf/cpm/Nun0KK20} used \emph{partitioning sets} \cite{DBLP:conf/soda/BirenzwigeGP20} to construct an anchor set of size $O(n/d)$. 
However, this latter anchor set takes too long time to construct to be used in our sublinear-time quantum algorithm.

Our key idea is to combine the oblivious approach and non-oblivious approach, and design anchor sets with a balance between the \emph{size} and the \emph{construction time}: the number of anchors is $m=O(n/d^{3/4})$, and, given any index $i\in [m]$, the $i^{\text{th}}$ anchor can be reported in $T=\tilde O(\sqrt{d})$ quantum time. 
Our construction (\cref{sec:anchor}) is based on an \emph{approximate version} of difference covers, combined with the \emph{string synchronizing sets} recently introduced by Kempa and Kociumaka \cite{bwt} (adapted to the sublinear setting using tools from pseudorandomness).
Roughly speaking, allowing errors in the difference cover makes the size much smaller, while also introducing slight misalignments between the anchors,  which are then to be fixed by the string synchronizing sets. 

\paragraph*{Anchoring via quantum walks.} Now we explain how to use small and explicit anchor sets to obtain better quantum LCS algorithms with time complexity $\tilde O(m^{2/3} \cdot (\sqrt{d}+T)) = \tilde O(n^{2/3})$, where $m = \tilde O(n/d^{3/4})$ is the number of anchors, and $T=\tilde O(\sqrt{d})$ is the time complexity of computing the $i^{\text{th}}$ anchor.  Our algorithm uses the \emph{MNRS quantum walk} framework \cite{mnrs} (see \cref{sec:qwalk}) on Johnson graphs. 
Informally speaking, to apply this framework, we need to solve the following dynamic problem: maintain a \emph{subset} of $r$ anchors which undergoes insertions and deletions (called \emph{update steps}), and in each query (called a \emph{checking step}) we need to solve the Two String Families LCP problem on this subset, i.e.,  answer whether the current subset contains a pair of anchors that can extend to a length-$d$ common substring. If each update step takes time $\mathsf{U}$, and each checking step takes time $\mathsf{C}$, then the MNRS quantum walk algorithm has overall running time $\tilde O(r \ \mathsf{U} + \frac{m}{r}\ (\sqrt{r} \ \mathsf{U} +\mathsf{C}))$. We will achieve $\mathsf{U} = \tilde O(\sqrt{d}+T)$ and $\mathsf{C} = \tilde O(\sqrt{rd})$, and obtain the claimed time complexity by setting $r = m^{2/3}$.

To solve this dynamic problem, we maintain the lexicographical ordering of the length-$d$ substrings specified by the current subset of anchors, as well as the corresponding LCP array which contains the length of the longest common prefix between every two lexicographically adjacent substrings. Note that the maintained information uniquely defines the \emph{compact trie} of these substrings. 
This information can be updated easily after each insertion (or deletion) operation: we first compute the inserted anchor in $T$ time, and then use binary search with Grover search to find its lexicographical rank and the LCP values with its neighbors, in $\tilde O(\sqrt{d})$ quantum time. 

The maintained information will be useful for the checking step. In fact, if we only care about query complexity, then we are already done, since the  maintained information already uniquely determines the answer of the Two String Families LCP problem, and no additional queries to the input strings are needed. The main challenge is to implement this checking step time-efficiently.
Unfortunately, the classical near-linear-time algorithm \cite{DBLP:conf/cpm/Charalampopoulos18} for solving the Two String Families LCP problem is too slow compared to our goal of $\sfC=\tilde O(\sqrt{rd})$, and it is not clear how to obtain a quantum speedup over this classical algorithm.
Hence, we should try to dynamically maintain the solution using data structures, instead of solving it from scratch every time.
In fact, such a data structure with $\polylog(n)$ time per operation was already given by Charalampopoulos, Gawrychowski, and Pokorski \cite{icalp20}, and was used to obtain a classical data structure for maintaining Longest Common Substring under character substitutions.
However, this data structure cannot be applied to the quantum walk algorithm, since it violates two requirements that are crucial for the correctness of quantum walk algorithms: 
(1) It should have worst-case time complexity (instead of being amortized), and 
(2) it should be \emph{history-independent} (see the discussion in \cref{sec:ds-overview} for more details). 
Instead, we will design a different data structure that satisfies these two requirements, and can solve the Two String Families LCP problem on the maintained subset in $\tilde O(\sqrt{rd})$ quantum time. 
This time complexity is worse than the $\polylog(n)$ time achieved by the classical data structure of \cite{icalp20}, but suffices for our application. 

\paragraph{A technical hurdle: limitations of 2D range query data structures.} Our solution for the Two String Families LCP problem is straightforward, but a key component in the algorithm relies on \emph{dynamic 2-dimensional orthogonal range queries}. This is a well-studied problem in the data structure literature,
and many $\polylog n$-time data structures are known (see \cite{range-search-old, full-dynamic-range, better-dynamic} and the references therein). 
However, for our results, the 2-dimensional (2D) range query data structure in question has to satisfy not only the two requirements mentioned above, but also a third requirement of being \emph{comparison-based}. 
In particular, we are not allowed to treat the coordinates of the 2D points as $\poly(n)$-bounded integers, because  the coordinates  actually  correspond to substrings of the input string, and should be compared by lexicographical order. 
Unfortunately, no data structures satisfying all three requirements are known.

To bypass this difficulty, our novel idea is to use a sampling procedure that lets us estimate  the rank of a coordinate of the inserted 2D point among all the possible coordinates, which effectively allows us to convert the non-integer coordinates into integer coordinates. 
By a version of the Balls-and-Bins hashing argument,  the inaccuracy incurred by the sampling  can be controlled for \emph{most} of the vertices on the Johnson graph which the quantum walk operates on.   
This then lets us apply 2D range query data structures over integer coordinates (see \cref{sec:apply-ds} for the details of this argument), which can be implemented with worst-case time complexity and history-independence as required. 
Combining this method with the tools and ideas mentioned before lets us get a time-efficient implementation of the quantum walk  algorithm for computing the LCS.

We believe this sampling idea will find further applications in improving the time efficiency of quantum walk algorithms (for example, it can simplify the implementation of Ambainis' $\tilde O(n^{2/3})$-time Element Distinctness algorithm, as noted in \cref{sec:open}).

\subsubsection{Minimal String Rotation}
In the Minimal String Rotation problem, we are given a string $s$ of length $n$ and are tasked with finding the cyclic rotation of $s$ which is lexicographically the smallest. 
We sketch the main ideas of our improved quantum algorithm for Minimal String Rotation by comparing it to the previous best solution for this problem.

The simplest version of Wang and Ying's algorithm  {\cite[Theorem 5.2]{ying}} works by identifying a small prefix of the minimal rotation using Grover search, and then applying pattern matching with this small prefix to find the starting position of the minimum rotation.
More concretely, let $B$ be some size parameter.
By quantum minimum finding over all prefixes of length $B$ among the rotations of $s$, we can find the length-$B$ prefix $P$ of the minimal rotation in asymptotically $\sqrt{B}\cdot \sqrt{n}$ time.
Next, split the string $s$ into $\Theta(n/B)$ blocks of size $\Theta(B)$ each. 
Within each block, we find the \emph{leftmost} occurrence of $P$ via quantum Exact String Matching \cite{matching}
It turns out that one of these positions is guaranteed to be a starting position of the minimal rotation (this property is called an ``exclusion rule'' or ``Ricochet Property'' in the literature).
By minimum finding over these $O(n/B)$ candidate starting positions (and comparisons of length-$n$ strings via Grover search), we can find the true minimum rotation in asymptotically $\sqrt{n/B} \cdot \sqrt{n}$ time.
So overall the algorithm takes asymptotically 
    \[\sqrt{Bn} + (n/\sqrt{B})\]
time, which is minimized at $B = \sqrt n$ and yields a runtime of $\tilde{O}(n^{3/4})$.

This algorithm is inefficient in its first step, where it uses quantum minimum finding to obtain the minimum length-$B$ prefix $P$. 
The length-$B$ prefixes we are searching over all  come from rotations of the same string $s$.
Due to this common structure, we should be able to find their minimum more efficiently than just using the generic algorithm for minimum finding.
At a high level, we improve this step by finding $P$ using \emph{recursion} instead.
Intuitively, this is possible because the Minimal Rotation problem is already about finding the minimum ``prefix'' (just of length $n$) among rotations of $s$.
This then yields a recursive algorithm running in $n^{1/2 + o(1)}$ quantum time.

In the presentation of this algorithm in \Cref{sec:lmsr}, we use a chain of reductions and actually solve a more general problem to get this recursion to work.
The argument also relies on a new ``exclusion rule,'' adapted from previous work, to prove that  we only need to consider a constant number of candidate starting positions of the minimum rotation within each small block of the input string.

\subsubsection{Longest Square Substring}
A \emph{square string} is a string of even length with the property that its first half is identical to its second half.
In other words, a string is square if it can be viewed as some string repeated twice in a row. 

We show how to find the longest square substring in an input string of length $n$ using a quantum algorithm which runs in $\tilde{O}(\sqrt n)$ time.
Our algorithm mostly follows the ideas used in \cite{legall} to solve the Longest Palindromic Substring problem, but makes some modifications due to the differing structures of square substrings and palindromic substrings (for example, \cite{legall} exploits the fact that if a string contains a large palindromic substring it has smaller palindromic substrings centered at the same position; in contrast, it is possible for a string to have a large square substring but not contain any smaller square substrings, so we cannot leverage this sort of property).

At a high level, our algorithm starts by guessing the size of the longest square substring within a $(1+\eps)$ factor for some small constant $\eps > 0$.
We then guess a large substring $P$ contained in the first half of an optimal solution, and then use the quantum algorithm for Exact String Matching to find a copy of this $P$ in the second half of the corresponding solution.
If we find a unique copy of $P$, we can use a Grover search to extend outwards from our copies of $P$ and recover a longest square substring.
Otherwise, if we find multiple copies, it implies our substring is periodic, so we can use a Grover search to find a maximal periodic substring containing a large square substring, and then employ some additional combinatorial arguments to recover the solution.

\subsection{Related Work}
\label{subsec:related}
\paragraph*{Quantum algorithms on string problems.}
Wang and Ying \cite{ying} improved the logarithmic factors of the quantum Exact String Matching algorithm by Ramesh and Vinay \cite{matching} (and filled in several gaps in their original proof), and showed that the same technique can be used to find the smallest period of a periodic string \cite[Appendix D]{ying}. 

Another important string problem is computing the \emph{edit distance} between two strings (the minimum number of deletions, insertions, and substitutions needed to turn one string into the other). The best known classical algorithm has $O(n^2/\log^2 n)$ time complexity \cite{MasekP80}, which is near-optimal under the Strong Exponential Time Hypothesis \cite{BackursI18}. 
It is open whether quantum algorithms can  compute edit distance in truly subquadratic time.
For the approximate version of the edit distance problem, the breakthrough work of Boroujeni et al.\ \cite{boroujeni2021approximating} gave a truly subquadratic time quantum algorithm for computing a constant factor approximation.
The quantum subroutines of this algorithm were subsequently replaced with classical randomized algorithms in
\cite{DBLP:journals/jacm/ChakrabortyDGKS20} to get a truly subquadratic classical algorithm that approximates the edit distance to a constant factor.

Le Gall and Seddighin \cite{legall} also considered the $(1+\eps)$-approximate Ulam distance problem (i.e., edit distance on non-repetitive strings), and showed a quantum algorithm with near-optimal $\tilde O(\sqrt{n})$ time complexity.  Their algorithm was based on the classical algorithm by Naumovitz, Saks, and  Seshadhri \cite{DBLP:conf/soda/NaumovitzSS17}.

Montarano  \cite{DBLP:journals/algorithmica/Montanaro17} gave quantum algorithms for the $d$-dimensional pattern matching problem with random inputs.
Ambainis et al.\ \cite{DBLP:conf/mfcs/AmbainisBIKKPSS20} gave quantum algorithms for deciding Dyck languages. There are also some results \cite{DBLP:journals/qic/AmbainisM14, DBLP:conf/swat/CleveIGNTTY12} on string problems with non-standard quantum queries to the input.

\paragraph*{Quantum walks and time-efficient quantum algorithms.}
\label{sec:intro-quantum}
Quantum walks \cite{DBLP:conf/focs/Szegedy04,DBLP:journals/siamcomp/Ambainis07,mnrs} are a useful method to obtain query-efficient quantum algorithms for  many important problems, such as
Element Distinctness \cite{DBLP:journals/siamcomp/Ambainis07}
and Triangle Finding  \cite{q-triangle,DBLP:conf/soda/JefferyKM13,DBLP:conf/focs/Gall14}. 
Ambainis showed that the query-efficient algorithm for element distinctness \cite{DBLP:journals/siamcomp/Ambainis07} can also be implemented in a time-efficient manner with only a $\polylog(n)$ blowup, by applying  history-independent data structures in the quantum walk. 
Since then, this ``quantum walk plus data structure'' strategy has been used in many quantum algorithms to obtain improved time complexity.  
For example, Belovs, Childs, Jeffery, Kothari, and Magniez \cite{DBLP:conf/icalp/BelovsCJKM13} used nested quantum walk with Ambainis' data structure to obtain time-efficient algorithms for the 3-distinctness problem. 
Bernstein, Jeffery, Lange, and Meurer \cite{DBLP:conf/pqcrypto/BernsteinJLM13} designed a simpler data structure called quantum radix tree \cite{jeffery2014frameworks}, and applied it in their quantum walk algorithms for the Subset Sum problem on random input. 
 Aaronson, Chia, Lin, Wang, and Zhang~\cite{ccc20aaronson} gave a quantum walk algorithm for the Closest-Pair problem in $O(1)$-dimensional space with near-optimal time complexity $\tilde O(n^{2/3})$.
 The previous  $\tilde O(n^{2/3})$-time algorithm   for approximating LCS in non-repetitive strings \cite{legall} also applied quantum walks.
 
 On the other hand, query-efficient quantum algorithms do not always have  time-efficient implementations.
 This motivated the study of \emph{quantum fine-grained complexity}. Aaronson et al.\ \cite{ccc20aaronson} formulated the QSETH conjecture, which is a quantum analogue of the classical Strong Exponential Time Hypothesis, and showed that Orthogonal Vectors and Closest-Pair in $\polylog(n)$-dimensional space require $n^{1-o(1)}$ quantum time under QSETH. In contrast, these two problems have simple quantum walk algorithms with only $O(n^{2/3})$ query complexity. Buhrman, Patro, and Speelman \cite{DBLP:conf/stacs/BuhrmanPS21} formulated another version of QSETH, which implies a conditional $\Omega(n^{1.5})$-time lower bound for quantum algorithms solving the edit distance problem.
 Recently, Buhrman, Loff, Patro, and Speelman \cite{DBLP:journals/corr/abs-2106-02005} proposed the quantum 3SUM hypothesis, and used it to show that the quadratic quantum speedups obtained by Ambainis and Larka \cite{AmbainisL20} for many computational geometry problems are conditionally optimal.
 Notably, in their fine-grained reductions, they employed a quantum walk with data structures  to bypass the linear-time preprocessing stage that a naive approach would require.

\paragraph*{Classical string algorithms.}
We refer readers to several excellent textbooks \cite{gusfield_1997,jewels,crochemore_hancart_lecroq_2007} on string algorithms.

Weiner \cite{DBLP:conf/focs/Weiner73} introduced the suffix tree and gave a linear-time algorithm for computing the LCS of two strings over a constant-sized alphabet.
For polynomially-bounded integer alphabets, Farach's construction of suffix trees \cite{DBLP:conf/focs/Farach97} implies an linear-time algorithm for LCS.
Babenko and Starikovskaya \cite{DBLP:conf/csr/BabenkoS08} gave an algorithm for LCS based on suffix arrays.
Recently, Charalampopoulos, Kociumaka, Pissis, and Radoszewski \cite{lcs2021} gave faster word-RAM algorithms for LCS  on compactly represented input strings over a small alphabet.
The LCS problem has also been studied in the settings of
time-space tradeoffs \cite{DBLP:conf/cpm/StarikovskayaV13,DBLP:conf/esa/KociumakaSV14,DBLP:conf/cpm/Nun0KK20}, approximate matching  \cite{DBLP:conf/csr/BabenkoS08, DBLP:conf/soda/AbboudWY15,DBLP:journals/ipl/FlouriGKU15,DBLP:journals/jcb/ThankachanAA16, DBLP:conf/cpm/Starikovskaya16, DBLP:journals/corr/abs-1712-08573,DBLP:conf/cpm/Charalampopoulos18, DBLP:conf/cpm/GourdelKRS20}, and dynamic data structures  \cite{DBLP:conf/esa/AmirCPR19,DBLP:journals/algorithmica/AmirCPR20,icalp20}.

Booth \cite{DBLP:journals/ipl/Booth80} and Shiloach \cite{DBLP:journals/jal/Shiloach81} gave the first linear time  algorithms for the Minimal String Rotation problem. Later, Duval \cite{DBLP:journals/jal/Duval83} gave a constant-space linear-time algorithm for computing the \emph{Lyndon factorization} of a string, which can be used to compute the minimal rotation, maximal suffix, and minimal suffix.
Duval's algorithm can also compute the minimal suffix and maximal suffix for every prefix of the input string. Apostolico and Crochemore \cite{DBLP:journals/iandc/ApostolicoC91} gave a linear-time algorithm for computing the minimal rotation of every prefix of the input string.
Parallel algorithms for Minimal String Rotation were given by Iliopoulos and Smyth \cite{DBLP:journals/tcs/IliopoulosS92}. There are data structures  \cite{DBLP:conf/cpm/BabenkoKS13,DBLP:journals/tcs/BabenkoGKKS16,DBLP:conf/cpm/Kociumaka16} that, given a substring specified by its position and length in the input string, can efficiently answer its minimal suffix, maximal suffix, and minimal rotation.
The Longest Lyndon Substring problem can be solved in linear time \cite{DBLP:journals/jal/Duval83} by simply outputting the longest segment in the Lyndon factorization. 
There are data structures \cite{DBLP:conf/cpm/UrabeNIBT18, DBLP:conf/esa/AmirCPR19} for dynamically maintaining the longest Lyndon substring  under character substitutions.

There are $O(n \log n)$-time algorithms for finding all the square substrings of the input string \cite{DBLP:journals/ipl/Crochemore81,DBLP:journals/jal/MainL84,DBLP:conf/paa/ApostolicoIP87}.
There are data structures \cite{DBLP:conf/esa/AmirBCK19} for dynamically maintaining the longest square substring under character substitutions.

The construction of difference cover \cite{DBLP:conf/cpm/BurkhardtK03,DBLP:journals/tocs/Maekawa85} has been previously used in many string algorithms, e.g.,  
\cite{DBLP:conf/cpm/BurkhardtK03,DBLP:conf/cpm/BilleGGLW15,DBLP:conf/cpm/GawrychowskiKRW16,lcs2021}. 
The string synchronizing set recently introduced by Kempa and Kociumaka \cite{bwt} has been applied in \cite{bwt,cpm19,lcs2021,DBLP:journals/corr/abs-2106-12725}. 
The local-consistency idea behind the construction of string synchronizing set had also appeared in previous work  \cite{DBLP:conf/soda/KociumakaRRW15, kociumaka, DBLP:conf/soda/BirenzwigeGP20}.

\subsection{Paper Organization}
In \Cref{sec:prelim}, we provide useful definitions and review some quantum primitives which will be used in our algorithms.
In \Cref{sec:lcs1}, we present our algorithm for \emph{Longest Common Substring}.
In \Cref{sec:lmsr}, we present our algorithm for \emph{Minimal String Rotation} and several related problems.
In \Cref{sec:squared}, we present our algorithm for \emph{Longest Square Substring}.
Finally, we mention several open problems in \Cref{sec:open}.

\section{Preliminaries}
\label{sec:prelim}

\subsection{Notations and Basic Properties of Strings}
\label{sec:prelim:notation}
We define sets $\N = \{0,1,2,3,\dots\}$ and $\N^+ = \{1,2,3,\dots\}$.
For every positive integer $n$ we introduce the set $[n]= \{1,2,\dots,n\}$. 
Given two integers $i\le j$,  we let $[i\dd j]= \{i,i+1,\dots,j\}$ denote the set of integers in the closed interval $[i,j]$.  
We define $[i\dd j),(i\dd j]$, and $(i\dd j)$ analogously. For an integer $x$ and an integer set $A$, let $x+A$ denote $\{x+a: a\in A\}$, and let $xA$ denote $\{x\cdot a: a\in A\}$.

%\cnote{TODO: change the terminology to be consistent with literature: a fragment is a pair of indices $i<j$. the fragment is an occurrence of the  underlying substring }

As is standard in the literature, we consider strings over a \emph{polynomially-bounded integer alphabet} $\Sigma = [1\dd |\Sigma|]$ of size $|\Sigma|\le n^{O(1)}$.
A string $s \in \Sigma^*$ of length $|s|=n$ is a sequence of characters $s=s[1]s[2]\cdots s[n]$ from the alphabet $\Sigma$ (we use 1-based indexing). 
The \emph{concatenation} of two strings $s,t\in \Sigma^*$ is denoted by $st$. The \emph{reversed string} of $s$ is denoted by  $s^{R}=s[n]s[n-1]\cdots s[1]$.

Given a string $s$ of length $|s|=n$, a \emph{substring} of $s$ is any string of the form $s[i\dd j] = s[i]s[i+1]\cdots s[j]$ for some indices $1\le i\le j\le n$. 
For $i>j$, we define $s[i\dd j]$ to be the empty string $\eps$. When $i,j$ may be out of bounds, we use the convention $s[i\dd j] = s[\max\{1,i\}\dd \min\{n, j\}]$ to simplify notation.
We sometimes use $s[i\dd j) = s[i]s[i+1]\cdots s[j-1]$ and $s(i \dd j]=s[i+1]\cdots s[j-1]s[j]$ to denote substrings. 
A substring $s[1\dd j]$ is called a \emph{prefix} of $s$, and a substring $s[i\dd n]$ is called a \emph{suffix} of $s$. 
For two strings $s,t$, let $\lcp(s,t)= \max\{j: j\le \min\{|s|,|t|\}, s[1\dd j] = t[1\dd j] \}$ denote the length of their \emph{longest common prefix}. 

We say string $s$ is \emph{lexicographically smaller} than string $t$ (denoted $s \prec t$) if either $s$ is a proper prefix of $t$ (i.e., $|s|<|t|$ and $s=t[1\dd |s|]$), or $\ell = \lcp(s,t)<\min\{|s|,|t|\}$ and $s[\ell+1]<t[\ell+1]$. The notations $\succ, \preceq,\succeq$ are defined analogously. The following easy-to-prove and well-known fact has been widely used in string data structures and algorithms.
\begin{lem}[e.g.\ {\cite[Lemma 1]{DBLP:journals/tcs/KentLS12}}]
\label[lemma]{lem:lcp-height}
Given strings $s_1\preceq s_2\preceq \cdots \preceq s_m$, we have $\lcp(s_1,s_m) = \min_{1\le i< m} \{ \lcp(s_i,s_{i+1})\}$.
\end{lem}

For a positive integer $p\le |s|$, we say $p$ is a \emph{period} of $s$ if $s[i] = s[i+p]$ holds for all $1\le i\le  |s|-p$. One can compute all the periods of $s$ by a classical deterministic algorithm in linear time \cite{DBLP:journals/siamcomp/KnuthMP77}. 
We refer to the minimal period of $s$ as \emph{the period} of $s$, and denote it by $\per(s)$.  
If $\per(s)\le |s|/2$, we say that $s$ is \emph{periodic}. If $\per(s)$ does not divide $|s|$, we say that $s$ is \emph{primitive}.  
We will need the following classical results regarding periods of strings for some of our algorithms.

\begin{lem}[Weak Periodicity Lemma, {\cite{periodicity}}]
\label[lemma]{weak-period}
If a string $s$ has periods $p$ and $q$ such
that $p + q \le |s|$, then $\gcd(p, q)$ is also a period of $s$.
\end{lem}

\begin{lem}[e.g., \cite{DBLP:conf/icalp/PlandowskiR98,DBLP:conf/soda/KociumakaRRW15}]
\label[lemma]{lm:period-patterns}
Let $s,t$ be two strings with $|s|/2 \le |t|\le |s|$, and let $s[i_1\dd i_1+|t|)= s[i_2\dd i_2+|t|)= \dots =  s[i_m\dd i_m+|t|) = t$ be all the occurrences of $t$ in $s$ (where $i_k<i_{k+1}$). Then, $i_1,i_2,\dots,i_m$ form an arithmetic progression. Moreover, if $m\ge 3$, then $\per(t) =i_2-i_1$.
\end{lem}

We say string $s$ is a \emph{(cyclic) rotation} of string $t$, if $|s|=|t|=n$ and there exists an index $1\le i \le n$ such that $s = t[i\dd n]t[1\dd i-1]$. If string $s$ is primitive and is lexicographically minimal among its cyclic rotations, we call $s$ a \emph{Lyndon word}.  Equivalently, $s$ is a Lyndon word if and only if $s\preceq t$ for all proper suffices $t$ of $s$. For a periodic string $s$ with minimal period $\per(s)=p$, the \emph{Lyndon root} of $s$ is defined as the lexicographically minimal rotation of $s[1\dd p]$, which can be computed by a classical deterministic algorithm in linear time (e.g., 
\cite{DBLP:journals/ipl/Booth80,DBLP:journals/jal/Shiloach81,DBLP:journals/jal/Duval83}).

\subsection{Problem Definitions}
\label{sec:problems}
We give formal definitions of the string problems considered in this paper.

\newcommand{\defproblem}[3]{
  \vspace{2mm}
%  \hline
  \vspace{1mm}
\noindent\fbox{
  \begin{minipage}{0.95\textwidth}
  #1 \\
  {\bf{Input:}} #2  \\
  {\bf{Task:}} #3
  \end{minipage}
  }
%  \vspace{1mm}
%  \hline
  \vspace{2mm}
}

\defproblem{Longest Common Substring}
{Two strings $s,t$}
{Output the maximum length $\ell$  such that
$s[i\dd i+\ell) = t[j\dd j+\ell)$ for some   $i \in [|s|-\ell+1], j \in [|t|-\ell+1]$.}

\defproblem{Longest Repeated Substring}
{A string $s$}
{Output the maximum length $\ell$  such that 
$s[i\dd i+\ell) = s[j\dd j+\ell)$
for some $i,j \in [|s|-\ell+1],i\neq j$.}

\defproblem{Longest Square Substring}
{A string $s$}
{Output the maximum shift $\Delta$ such that $s[i\dd i+\Delta)=s[i+\Delta\dd i+2\Delta)$ for some  $i\in [1\dd |s|-2\Delta+1]$.
}

\defproblem{Longest Lyndon Substring}
{A string $s$}
{Output the maximum length $\ell$ such that $s[i\dd i+\ell)$ is a Lyndon word for some $i\in [|s|-\ell+1]$.
}

\defproblem{Exact String Matching}
{Two strings $s,t$ with $|s|\ge |t|$}
{Output the minimum position $i$  such that
$s[i\dd i+|t|) = t$.}

\defproblem{Minimal String Rotation}
{A string $s$}
{Output a position $i\in [1\dd |s|]$ such that $s[i\dd |s|]s[1\dd i-1]\preceq s[j\dd |s|]s[1\dd j-1]$ holds for all $j\in [1\dd |s|]$. If there are multiple solutions, output the smallest such $i$. }

\defproblem{Maximal Suffix}
{A string $s$}
{Output the position $i\in [1\dd |s|]$ such that $s[i\dd |s|]\succ s[j\dd |s|]$ holds for all $j\in [|s|] \setminus \{i\}$. }

\defproblem{Minimal Suffix}
{A string $s$}
{Output the position $i\in [1\dd |s|]$ such that $s[i\dd |s|]\prec s[j\dd |s|]$ holds for all $j\in [|s|] \setminus \{i\}$. }

In the first four problems, we only require the algorithm to output the maximum length. The locations of the witness substrings can be found by a binary search.

\subsection{Computational Model}
\label{sec:model}
We assume the input strings can be accessed in a quantum query model \cite{ambainis2004quantum,DBLP:journals/tcs/BuhrmanW02}, which is standard in the literature of quantum algorithms.  More precisely, letting $s$ be an input string of length $n$, we have access to an oracle $O_s$ that, for any index $i\in [n]$ and any $b\in \Sigma$, performs the unitary mapping $O_s\colon \ket {i,b}  \mapsto \ket{i,b\oplus s[i]}$, where $\oplus$ denotes the XOR operation on the binary encodings of characters. The oracles can be queried in superposition, and each query has unit cost. Besides the input queries, the algorithm can also apply intermediate unitary operators that are independent of the input oracles.
Finally, the query algorithm should return the correct answer with success probability at least $2/3$ (which can be boosted to high probability\footnote{We say an algorithm succeeds \emph{with high probability (w.h.p)}, if the success probability can be made at least $1-1/n^c$ for any desired constant $c>1$.} by a majority vote over $O(\log n)$ repetitions). 
The \emph{query complexity} of an algorithm is the number of queries it makes to the input oracles.

In this paper, we are also interested in the \emph{time complexity} of the quantum algorithms, which counts not only the queries to the input oracles, but also the  elementary gates \cite{PhysRevA.52.3457} for implementing the unitary operators that are independent of the input. In order to implement the query algorithms in a time-efficient manner, we also need the \emph{quantum random access gate}, defined as 
\[ \ket{i,b,z_1,\dots,z_m} \mapsto \ket{i,z_i,z_1,\dots,z_{i-1},b,z_{i+1},\dots,z_m},\]
to access at unit cost the $i^{\text{th}}$ element from the quantum working memory $\ket{z_1,\dots,z_m}$.
Assuming quantum random access, a classical time-$T$ algorithm that uses random access memory can be converted into a quantum subroutine in time $O(T)$, which can be invoked by quantum search primitives such as Grover search.
Quantum random access has become a standard assumption in designing time-efficient quantum algorithms (for example, all the time-efficient quantum walk algorithms mentioned in \cref{sec:intro-quantum} relied on this assumption). 

\subsection{Basic Quantum Primitives}
\label{sec:primitive}

\paragraph*{Grover search (Amplitude amplification) \cite{DBLP:conf/stoc/Grover96,brassard2002quantum}.}
Let $f\colon [n] \to \{0,1\}$ be a function, where $f(i)$ for each $i\in [n]$ can be evaluated in time $T$. There is a quantum algorithm that, with high probability, finds an $x\in f^{-1}(1)$ or report that $f^{-1}(1)$ is empty, in $\tilde O(\sqrt{n}\cdot T)$ time.  Moreover, if it is guaranteed that either $|f^{-1}(1)| \ge M$ or $|f^{-1}(1)|=0$ holds, then the algorithm runs in $\tilde O(\sqrt{n/M} \cdot T)$ time.

\paragraph*{Quantum minimum finding \cite{minimumfinding}.} Let $x_1,\dots,x_n$ be $n$ items with a total order, where each pair of $x_i$ and $x_j$ can be compared in time $T$. There is a quantum algorithm that, with high probability, finds the minimum item among $x_1,\dots,x_n$ in $\tilde O(\sqrt{n} \cdot T)$ time.

\begin{remark}
If the algorithm for evaluating $f(i)$ (or for comparing $x_i,x_j$) has some small probability of outputting the wrong answer, we can first boost it to high success probability, and then the Grover search (or Quantum minimum finding) still works, since quantum computational errors only accumulate linearly.
It is possible to improve the log-factors in the query complexity of quantum search when the input has errors \cite{DBLP:conf/icalp/HoyerMW03}, but in this paper we do not seek to optimize the log-factors.
\end{remark}

\begin{lem}[Computing LCP]
\label[lemma]{lem:lcp-grover}
Given two strings $s,t$ of lengths $|s|,|t|\le n$, there is a quantum algorithm that computes $\lcp(s,t)$ and decides whether $s\preceq t$, in $\tilde O(\sqrt{n})$ time.
\end{lem}
\begin{proof}
Note that we can use Grover search to decide whether two strings are identical in $\tilde O(\sqrt{n})$ time. Then we can compute $\lcp(s,t)$ by a simple binary search over the length of the prefix. After that we can easily compare their lexicographical order by comparing the next position.
\end{proof}

Given a string $s$ and positions $g$ and $h$ such that $s[g] = s[h]$, we often use \Cref{lem:lcp-grover} to ``extend'' these common characters to larger identical strings to some bound $d$ while keeping them equivalent (i.e. find the largest positive integer $j\le d$ such that $s[g\dd g+j) = s[h\dd h+j)$).
We will often refer to this process (somewhat informally) as ``extending strings via Grover search.''

As a final useful subroutine, we appeal to the result of
Ramesh and Vinay \cite{matching}, who combined Grover search with the deterministic sampling technique of Vishkin \cite{DBLP:journals/siamcomp/Vishkin91}, and obtained a quantum algorithm for Exact String Matching.
\begin{thm}[Quantum Exact String Matching \cite{matching}]
We can solve the Exact String Matching problem with a quantum algorithm on input strings $s,t$ of length at most $n$  using $\tilde{O}(\sqrt{n})$ query complexity and time complexity.
\end{thm}

\subsection{Quantum Walks}
\label{sec:qwalk}
We use the quantum walk framework developed by Magniez, Nayak, Roland, and Santha \cite{mnrs}, and apply it on Johnson graphs,

The Johnson graph $J(m,r)$ has $\binom{m}{r}$ vertices, each being an subset of $[m]$ with size $r$, where two vertices in the graph $A,B\in \binom{[m]}{r}$ are connected by an edge if and only if $|A\cap B|  = r-1$, or equivalently there exist $a\in A, b\in [m]\setminus A$ such that $B = (A \setminus \{a\}) \cup \{b\}$.  Depending on the application, we usually identify  a special  subset of the vertices $V_{\textsf{marked}}\subseteq \binom{[m]}{r}$ as being 
\emph{marked}. 
The quantum walk is analogous to a random walk on the Johnson graph attempting to find a marked vertex, but provides quantum speed-up compared to the classical random walk. The vertices in the Johnson graph are also called the states of the walk.

In the quantum walk algorithm, each vertex $K\in \binom{[m]}{r}$ is associated with a data structure $D(K)$. The setup cost $\sfS$ is the cost to set up the data structure $D(K)$ for any $K\in \binom{[m]}{r}$, where the cost could be measured in query complexity or time complexity. The checking cost $\sfC$ is the cost to check whether $K$ is a marked vertex, given the data structure $D(K)$. The update cost $\sfU$ is the cost of updating the data structure from $D(K)$ to $D(K')$, where $K' = (K\setminus \{a\})\cup \{b\}$ is an adjacent vertex specified by $a\in K, b\in [m]\setminus K$. The MNRS quantum walk algorithm  can be summarized as follows.
\begin{thm}[MNRS Quantum Walk \cite{mnrs}]
\label{thm:mnrs}
Suppose $|V_{\sf{marked}}|/\binom{m}{r} \ge \eps$ whenever $V_{\sf{marked}}$ is non-empty. Then there is a quantum algorithm that with high probability determines if $V_{\sf{marked}}$ is empty or finds a marked vertex, with cost of order $\sfS + \frac{1}{\sqrt{\eps}} ( \sqrt{r}\cdot \sfU + \sfC )$.
\end{thm}

Readers unfamiliar with the quantum walk approach are referred to \cite[Section 8.3.2]{notes} for a quick application of this theorem to solve the Element Distinctness problem using  $O(n^{2/3})$ quantum queries. This algorithm can be implemented in $\tilde O(n^{2/3})$ time by carefully designing the data structures to support time-efficient insertion, deletion, and searching \cite[Section 6.2]{DBLP:journals/siamcomp/Ambainis07}. We elaborate on the issue of time efficiency when we apply quantum walks in our algorithm in \cref{sec:lcs2}.

\section{Longest Common Substring}
\label{sec:lcs1}
In this section, we prove the following theorem.
\begin{thm}
\label{thm:lcs-main}
The Longest Common Substring (LCS) problem can be solved by a quantum algorithm with $\tilde O(n^{2/3})$ query complexity and time complexity.
\end{thm}

In \Cref{sec:lcs-warm-up}, we will give an outline of our quantum walk algorithm based on the notion of \emph{good anchor sets}, and show that this algorithm achieves good \emph{query complexity}.
Then in \Cref{sec:lcs2}, we describe how to use data structures to implement the quantum walk algorithm in a time-efficient manner. 
Finally, in \Cref{sec:anchor}, we present the construction of good anchor sets used by the algorithm.
We have organized the arguments so that
\Cref{sec:lcs2} and  \Cref{sec:anchor} are independent of one another, and can be read separately.

\subsection{Anchoring via Quantum Walks}
\label{sec:lcs-warm-up}

As mentioned in \cref{sec:overview-lcs}, our algorithm for LCS is based on the anchoring technique which previously appeared in classical LCS algorithms. Here, we will implement this technique using the MNRS quantum walk framework (\cref{sec:qwalk}). 

\paragraph*{Notations and input assumptions.} To simplify the presentation, we concatenate the two input strings $s,t$ into $S:=s\$ t$, where $\$$ is a delimiter symbol that does not appear in the input strings, and let $n = |S| =|s|+1+|t|$. So $s[i]=S[i]$ for all $i\in [1\dd |s|]$, and $t[j] = S[|s|+1+j]$ for all $j\in [1\dd |t|]$. 

We will only solve the \emph{decision version} of LCS: given a length threshold $d$,  determine whether $s$ and $t$ have a common substring of length $d$. 
The algorithm for computing the length of the longest common substring then follows from a binary search over the threshold $d$. 
We assume $d\ge 100$ to avoid corner cases in later analysis; for smaller $d$, the problem can be solved in $\tilde O(n^{2/3}d^{1/2}) = \tilde O(n^{2/3})$ time by reducing to the (bipartite version of) element distinctness problem  \cite[Section 3.1.1]{legall} and applying Ambainis' algorithm \cite{DBLP:journals/siamcomp/Ambainis07} (see \cref{sec:overview-lcs}).

\paragraph*{Anchoring.}
We begin by introducing the notion of \emph{good anchor sets}.
\begin{definition}[Good anchor sets]
	\label[definition]{defn:good-anchor}
For input strings $s,t$ and threshold length $d$, we call $C\subseteq [1\dd n]$ a \emph{good anchor set} if the following holds: if the longest common substring of $s$ and $t$ has length at least $d$, then there exist positions $i \in [1\dd |s|-d+1],j\in [1\dd |t|-d+1]$ and a shift $h\in [0\dd d)$, such that $s[i\dd i+d)=t[j\dd j+d)$, and $i+h,  |s|+1+j+h \in C$. 
	
In this definition, the anchor set $C$ is allowed to depend on $s$ and $t$.
If $C = \{C(1),C(2),\dots,C(m)\}$ and there is a (quantum) algorithm that, given any index $1\le j\le m$, computes the element $C(j)$ in $T(n,d)$ time, then we say $C$ is \emph{$T(n,d)$-(quantum)-time constructible}. 
The elements $C(1),C(2),\dots,C(m)$ are allowed to contain duplicates (i.e., $C$ could be a multiset), and are not necessarily sorted in any particular order.
\end{definition}

The set $[1\dd n]$ is trivially a good anchor set, but there are constructions of much smaller size. As a concrete example, one can directly construct good anchor sets using  \emph{difference covers}.
\begin{definition}[Difference cover  \cite{DBLP:conf/cpm/BurkhardtK03,DBLP:journals/tocs/Maekawa85}]
\label[definition]{defn:dlcover-easy}
A set $D\subseteq \N^+$ is called a \emph{$d$-cover}, if for every $i,j\in \N^+$, there exists an integer $h(i,j) \in [0\dd d)$ such that $i+h(i,j),j+h(i,j) \in D$.
\end{definition}
The following construction of $d$-cover has optimal size (up to a constant factor).
\begin{lem}[Construction of $d$-cover \cite{DBLP:conf/cpm/BurkhardtK03,DBLP:journals/tocs/Maekawa85}]
\label[lemma]{lem:construct-dcover-easy}
For every positive integer $d\ge 1$, there is a $d$-cover $D$ such that $D\cap [n]$ contains $O(n/\sqrt{d})$ elements. 
Moreover, given integer $i\ge 1$, one can compute the $i^{\text{th}}$ smallest element of $D\cap [n]$ in $\tilde O (1)$ time.
\end{lem}
Here we omit the proof of \Cref{lem:construct-dcover-easy}, as a more general version (\Cref{lem:construct-dcover}) will be proved later in \Cref{sec:diffcover}.
Using difference covers, we immediately have the following simple construction of good anchor sets.
\begin{cor}[A simple good anchor set]
\label[corollary]{cor:good-anchor-easy}
 There is a $\tilde O(1)$-time constructible  good anchor set $C$ of size $m= O(n/\sqrt{d})$.
\end{cor}
\begin{proof}
Let $D$ be the $d$-cover from \Cref{lem:construct-dcover-easy}. Then, for input strings $s,t$ and threshold length $d$, it immediately follows from definition that $C:=\big (D\cap [|s|]\big ) \cup \big (|s|+1+(D\cap [|t|])\big )$ is a good anchor set.
\end{proof}
Note that the construction in \Cref{cor:good-anchor-easy} is deterministic, and oblivious to the content of the input strings $s$ and $t$.
The following lemma (which will be proved in \Cref{sec:anchor}) states that we can achieve a smaller size by a probabilistic non-oblivious construction that takes longer time to compute.   
\begin{lem}[A smaller good anchor set]
	\label[lemma]{lem:good-anchor-}
	There is an $\tilde O(\sqrt{d})$-quantum-time constructible anchor set $C$ of size $m= O(n/d^{3/4})$. This set $C$ depends on the input strings $s,t$ and $O(\log n)$ many random coins, and is a good anchor set with at least $2/3$ probability over the random coins.
\end{lem}

Let $C=\{C(1),\dots,C(m)\} \subseteq[n]$ be a good anchor set of size $|C|=m$. For every anchor $C(k)$ indexed by $k\in [m]$, we associate it with a pair of strings $(P(k),Q(k))$, where
\begin{align*}
    P(k)&:=S [C(k)\dd C(k)+d ),\\Q(k)&:=\big (S(C(k)-d\dd C(k)-1]\big )^R
\end{align*} are substrings (or reversed substrings) of $S$ obtained by extending from the anchor $C(k)$ to the right or reversely to the left.
The length of $P(k)$ is at most\footnote{Recall that we use the convention $S[x\dd y) := S[\max\{1,x\}\dd \min\{y+d,n+1\})$ for a length-$n$ string $S$.} $d$, and the length of $Q(k)$ is at most $d-1$.
We say the string pair $(P(k),Q(k))$ is \emph{red} if $C(k)\in [1\dd |s|]$, or \emph{blue} if $C(k)\in [|s|+1\dd n]$. We also say $k\in [m]$ is a \emph{red index} or a \emph{blue index}, depending on the color of the string pair $(P(k),Q(k))$.
Then, from the definition of good anchor sets, we immediately have the following simple observation.
\begin{prop}[Witness Pair]
\label[proposition]{prop:witness-pair}
The longest common substring of $s$ and $t$ has length at least $d$, if and only if there exist a red string pair $(P(k),Q(k))$ and a blue string pair $(P(k'),Q(k'))$ where $k,k'\in [m]$, such that $\lcp(P(k),P(k')) + \lcp(Q(k),Q(k')) \ge d$. In such case, $(k,k')$ is called a \emph{witness pair}.
\end{prop}
\begin{proof}
Suppose $s$ and $t$ have LCS of length at least $d$. Then the property of the good anchor set $C$ implies the existence of a shift $h\in [0\dd d)$ and a length-$d$ common substring $s[i\dd i+d)=t[j\dd j+d)$ such that $i+h=C(k),  |s|+1+j+h = C(k')$ for some $k,k'\in [m]$. Then, we must have $\lcp(P(k),P(k'))\ge d-h$ and $\lcp(Q(k),Q(k')) \ge h$, implying that $(k,k')$ is a witness pair.

Conversely, the existence of a witness pair immediately implies a common substring of length at least $d$.
\end{proof}

\begin{remark}
\label[remark]{remark:repeat}
The algorithm we are going to describe can be easily adapted to the Longest Repeated Substring problem: we only have one input string $S[1\dd n]$, and we drop the red-blue constraint in the definition of witness pairs in \cref{prop:witness-pair}.
\end{remark}
Now we shall describe our quantum walk algorithm that solves the decision version of LCS by searching for a witness pair.

\paragraph*{Definition of the Johnson graph.}
Recall that $C=\{C(1),\dots,C(m)\}$ is a good anchor set of size $|C|=m$.
We perform a quantum walk on the Johnson graph with vertex set $\binom{[m]}{r}$,  where $r$ is a parameter to be determined later. 
A vertex $K=\{k_{1},k_2,\dots,k_r\}\subseteq [m]$ in the Johnson graph is called a \emph{marked vertex}, if and only if $\{k_{1},k_2,\dots,k_r\}$ contains a witness pair (\Cref{prop:witness-pair}). If $s$ and $t$ have a common substring of length $d$,  then  at least $\binom{m-2}{r-2} / \binom{m}{r} = \Omega(r^2/m^2)$ fraction of the vertices are marked. Otherwise, there are no marked vertices.

\paragraph*{Associated data.}
In the quantum walk algorithm, each state $K=\{k_1,\dots,k_r\}\subseteq [m]$ is associated with the following data.
\begin{itemize}
    \item The indices $k_1,\dots,k_r$ themselves.
    \item The corresponding anchors $C(k_1),\dots,C(k_r) \in [n]$.
    \item An array $(k^P_1,\dots,k^P_r)$, which is a permutation of $k_1,\dots,k_r$, such that $P(k_i^P)\preceq P(k_{i+1}^P)$ for all $1\le i<r$.
    \item The LCP array $h_1^P,\dots,h^P_{r-1}$, where  $h_i^P = \lcp(P(k_i^P),P(k_{i+1}^P))$ 
    \item An array $(k^{Q}_1,\dots,k^{Q}_r)$, which is a permutation of $k_1,\dots,k_r$, such that $Q(k_i^{Q})\preceq Q(k_{i+1}^{Q})$ for all $1\le i<r$.
    \item The LCP array $h_1^Q,\dots,h_{r-1}^Q$, where  $h_i^Q = \lcp(Q(k_i^{Q}),Q(k_{i+1}^{Q}))$.
\end{itemize}

Note that we stored the \emph{lexicographical orderings} of the strings $P(k_1),\dots,P(k_r)$ and $Q(k_1),\dots,Q(k_r)$ (for identical substrings, we break ties by comparing the indices themselves), as well  as  the \emph{LCP arrays} which include the length of the longest common prefix of every pair of lexicographically adjacent substrings.
 By \Cref{lem:lcp-height}, these arrays together uniquely determine the values of $\lcp\big (P(k_i),P(k_j)\big )$ and $\lcp\big (Q(k_i),Q(k_j)\big )$, for \emph{every} pair of $i,j\in [r]$.\footnote{To better understand this fact, observe that they uniquely determine the \emph{compact tries} of $P(k_1),\dots,P(k_r)$ and of $Q(k_1),\dots,Q(k_r)$, where the LCP of two strings equals the depth of the lowest common ancestor of the corresponding nodes in the compact trie.}

In the checking step of the quantum walk algorithm, we decide whether the state is marked, by searching for a witness pair (\cref{prop:witness-pair}) in $\{k_1,\dots,k_r\}$.
Note that the contents of the involved strings $\{P(k_i)\}_{i\in [r]}$, $\{Q(k_i)\}_{i\in [r]}$ are no longer needed in order to solve this task, as long as we already know their lexicographical orderings and the LCP arrays.  This task is termed as the \emph{Two String Families LCP} problem in the literature \cite{DBLP:conf/cpm/Charalampopoulos18}, formalized as below.

\defproblem{Two String Families LCP}
{$r$ red/blue pairs of strings $(P_1,Q_1),(P_2,Q_2),\dots,(P_r,Q_r)$ of lengths $|P_i|,|Q_i|\le d$, which are represented by the lexicographical orderings of $P_1,\dots,P_r$ and of $Q_1,\dots,Q_r$, and their LCP arrays}
{Decide if there exist a red pair $(P,Q)$ and a blue pair $(P',Q')$, such that $\lcp(P,P')+\lcp(Q,Q') \ge d$. }

We will show how to solve this task time-efficiently in \cref{sec:lcs2}. 
For now, we only consider the query complexity of the algorithm, and we have the following simple observation, due to the fact that our associated information already uniquely determines the LCP value of every pair.
\begin{prop}[Query complexity of checking step is zero]
\label[proposition]{prop:checking-query}
Using the associated data defined above, we can determine whether $\{k_1,\dots,k_r\}\subseteq [m]$ is a marked state, without making any additional queries to the input.
\end{prop}

Now, we consider the cost of maintaining the associated data when the subset $\{k_1,\dots,k_r\}$ undergoes insertion and deletion during the quantum walk algorithm. 
\begin{prop}[Update cost]
\label{prop:update-query-cost}
Assume the anchor set $C$ is $T$-time constructible. Then, each update step of the quantum walk algorithm has query complexity $ \sfU = \tilde O(\sqrt{d} + T)$.
\end{prop}
\begin{proof}
Let us consider how to update the associated data when a new index $k$ is being inserted into the subset $\{k_1,\dots,k_r\}$. The deletion process is simply the reverse operation of insertion.

The insertion procedure can be summarized by the pseudocode in Algorithm~\ref{algo:insert}.
	First, we compute and store $C(k)$ in time $T$. Then we use a binary search to find the correct place to insert $k$ into the lexicographical orderings $(k_1^P,\dots,k_r^P)$ (and $(k_1^Q,\dots,k_r^Q)$).  Since the involved substrings have length $d$, each lexicographical comparison required by this binary search can be implemented in  $\tilde O(\sqrt{d})$ time by \Cref{lem:lcp-grover}.
	After inserting $k$ into the list, we update the LCP array by computing its LCP values $h_{\mathsf{pre}},h_{\mathsf{suc}}$ with two neighboring substrings, and removing (by ``uncomputing'')  the LCP value $h_{\sf{old}}$ between their neighbors which were adjacent at first, in $\tilde O(\sqrt{d})$ time (\Cref{lem:lcp-grover}).   \end{proof}
\begin{algorithm2e}
	Given an index $k \in [m]$\\
	Compute $C(k)$ \label{line:computeck}\\
	Store the data $(k,C(k))$ \\
	Compute the rank $i$ of $P(k)$ among $P(k_1^P),\dots,P(k_r^P)$ \label{line:start}\\
	Compute $h_{\textsf{pre}} = \lcp(P(k_{i-1}^P),P(k))$\\
	Compute $h_{\textsf{suc}} = \lcp(P(k_{i}^P),P(k))$\\
	Compute $h_{\textsf{old}} = \lcp(P(k_{i-1}^P),P(k_{i}^P))$ \label{line:old}\\
	Update $(k_1^P,\dots,k_r^P) \gets (k_1^P,\dots,k_{i-1}^P,k,k_{i}^P,\dots,k_r^P)$ \label{line:updk}\\
	Update $(h_1^P,\dots,h_{r-1}^P) \gets (h_1^P,\dots,h_{i-2}^P,h_{\textsf{pre}},h_{\textsf{suc}},h_{i}^P,\dots,h_{r-1}^P)$ \label{line:finish}\\
	Update $(k_1^{Q},\dots,k_r^{Q})$ and $(h_1^Q,\dots,h_{r-1}^Q)$ similarly as in Lines \ref{line:start}-\ref{line:finish}\\
	\caption{The insertion procedure}
	\label{algo:insert}
\end{algorithm2e}

\begin{prop}[Setup cost]
The  setup step of the quantum walk has query complexity $\sfS = \tilde O(r\cdot (\sqrt{d}+T))$.
\end{prop}
\begin{proof}
   We can set up the initial state for the quantum walk by simply performing $r$ insertions successively using \Cref{prop:update-query-cost}. 
\end{proof}

\begin{remark}
\label[remark]{remark:insertiontime}
Observe that, in the insertion procedure in Algorithm~\ref{algo:insert}, Lines \ref{line:computeck} and \ref{line:start}-\ref{line:old} can be implemented also in \emph{time complexity} $\tilde O(\sqrt{d}+T)$. 
The time-consuming steps in Algorithm~\ref{algo:insert} are those that actually modify the data. 
For example, in Lines \ref{line:updk} and \ref{line:finish}, the insertion causes some elements in the array to shift to the right, and would take $O(r)$ time if implemented naively.
Later in \Cref{sec:lcs2} we will describe appropriate data structures to implement these steps time-efficiently. 
\end{remark}

Finally, by \Cref{thm:mnrs}, the query complexity of our quantum walk algorithm  (omitting $\polylog(n)$ factors) is
\begin{align}
&    \sfS + \sqrt{\frac{m^2}{r^2}}\cdot (\sfC + \sqrt{r} \cdot \sfU) \label{eqn:cxty}\\
	= \ & r\cdot (\sqrt{d}+T) + \frac{m}{r}\cdot \big (0 + \sqrt{r}\cdot (\sqrt{d}+T)\big )\nonumber\\
	= \ & m^{2/3} \cdot (\sqrt{d} + T),\nonumber
\end{align}
by choosing $r = m^{2/3}$.
 The construction of good anchor sets from \Cref{cor:good-anchor-easy} has $m = O(n/\sqrt{d}), T=\tilde O (1)$, achieving query complexity $\tilde O(n^{2/3}\cdot d^{1/6})$.
The improved construction from \Cref{lem:good-anchor-} has $m = O(n/d^{3/4}), T =\tilde O(\sqrt{d})$, achieving query complexity $\tilde O(n^{2/3})$.

\subsection{Time-efficient Implementation}
\label{sec:lcs2}
In this section, we will show how to implement the $\tilde O(n^{2/3})$-query quantum walk algorithm from \cref{sec:lcs-warm-up} in \emph{time complexity} $\tilde O(n^{2/3})$. 

\subsubsection{Overview}
\label{sec:ds-overview}
Recall that our algorithm described in \Cref{sec:lcs-warm-up} for input strings $s,t$ and threshold length $d$ performs a quantum walk on the Johnson graph $\binom{[m]}{r}$. 
In this section, we have to measure the quantum walk costs $\sfS,\sfC,\sfU$ in terms of the \emph{time complexity} instead of query complexity. 
Inspecting \Cref{eqn:cxty}, we observe that the quantum walk algorithm can achieve $\tilde O(n^{2/3})$ time complexity, as long as we can implement the setup, checking and update steps with time complexities $\sfS = \tilde O(r(\sqrt{d}+T)), \sfC = \tilde O(\sqrt{rd})$, and $\sfU = \tilde O(\sqrt{d}+T) $.

As mentioned in \cref{sec:lcs-warm-up}, there are two parts in the described quantum walk algorithm that are time-assuming:
\begin{itemize}
    \item Maintaining the arrays of associated data under insertions and deletions (\cref{remark:insertiontime}).
    \item Solving the Two String Families LCP problem in the checking step.
\end{itemize}
Now we give an overview of how we address these two problems.
\paragraph*{Dynamic arrays under insertions and deletions.}
A natural solution to speed up the insertions and deletions is to maintain the arrays of using  appropriate data structures, which support the required operations in $\tilde O(1)$ time.  
This ``quantum walk plus data structures'' framework was first used in Ambainis' element distinctness algorithm \cite{DBLP:journals/siamcomp/Ambainis07}, and have been applied to many time-efficient quantum walk algorithms (see the discussion in \cref{subsec:related}).
However, as noticed by Ambainis \cite[Section 6.2]{DBLP:journals/siamcomp/Ambainis07}, such data structures have to satisfy the following requirements in order to be applicable in quantum walk algorithms.
\begin{enumerate}
	\item The data structure needs to be \emph{history-independent}, that is, the representation of the data structure in memory should only depend on the set of elements stored (and the random coins used) by the data structure, \emph{not} on the sequence of operations leading to this set of elements.
	\label{item-history}
	\item The data structure should guarantee  \emph{worst-case} time complexity (with high probability over the random coins) per operation. \label{item-worst}
\end{enumerate}

The first requirement guarantees that each vertex of the Johnson graph corresponds to a unique quantum state, which is necessary since having multiple possible states would destroy the interference during the quantum walk algorithm. This requirement rules out many types of self-balancing binary search trees\footnote{One exception is Treap.} such as AVL Tree and Red-Black Tree. 

The second requirement rules out data structures with amortized or expected running time, which may take very long time in some of the operations. The reason is that, during the quantum algorithm, each operation is actually applied to a superposition of many instances of the data structure, so we would like the time complexity of an operation to have a fixed upper bound that is independent of the particular instance being operated on.

Ambainis \cite{DBLP:journals/siamcomp/Ambainis07} designed a data structure satisfying both requirements based on hash tables and skip lists, which maintains a sorted list of items, and supports insertions, deletions, and searching in $\tilde O(1)$ time with high probability. 
Buhrman, Loff, Patro, and Speelman \cite{DBLP:journals/corr/abs-2106-02005} modified this data structure to also support indexing queries,  which ask for the $k^{\text{th}}$ item 
in the current list (see \cref{lem:skiplist} below). 
Using this data structure to maintain the arrays in our quantum walk algorithm, we can implement the update steps and the setup steps time-efficiently.

\paragraph*{Dynamic Two String Families LCP.} The checking step of our quantum walk algorithm (\Cref{prop:checking-query}) requires solving an \emph{Two String Families LCP} instance with $r$ string pairs of lengths bounded by $d$. 
We will not try to solve this problem from scratch for each instance, since it is not clear how to solve it significantly faster than the $\tilde O(r)$-time classical algorithm  \cite[Lemma 3]{DBLP:conf/cpm/Charalampopoulos18} even using quantum algorithms.
Instead, we \emph{dynamically maintain} the solution using some data structure, which efficiently handles each update step during the quantum walk where we insert one string pair $(P,Q)$ into (and remove one from) the current Two String Families LCP instance. 
As mentioned in \cref{sec:overview-lcs}, the classical data structure for this task given by Charalampopoulos, Gawrychowski, and Pokorski \cite{icalp20} is not applicable here, since it violates both requirements mentioned above: it maintains a heavy-light decomposition of the compact tries of the input strings, and rebuilds them from time to time to ensure amortized $\polylog(n)$ time complexity. 
It is not clear how to implement this strategy in a history-independent way and with worst-case time complexity per operation.

Instead, we will design a different data structure that satisfies the history-independence and worst-case update time requirements, and can solve the Two String Families LCP problem on the maintained instance in $\tilde O(\sqrt{rd})$ quantum time. 
This time complexity is much worse than the $\polylog(n)$ time achieved by the classical data structure of \cite{icalp20}, but is sufficient for our purpose. 
As mentioned in \cref{sec:overview-lcs}, one challenge is the lack of a \emph{comparison-based} data structure for 2D range query that also satisfies the two requirements above. We remark that there  exist comparison-based data structures with history-independence but only with expected time complexity  (e.g., \cite{DBLP:conf/swat/BlellochGV08}). There also exist folklore data structures for \emph{integer coordinates} that have history-independence and worst-case time complexity (e.g., \cref{lem:range}).
For the easier problem of 1-dimensional range query, there exist folklore data structures (e.g., \cref{lem:skiplist}) that satisfy all three requirements. 
To get around this issue, we will use a sampling procedure and a version of the Balls-and-Bins argument, which can effectively convert the involved non-integer coordinates into integer coordinates. Then, we are able to apply 2D range query data structures over integer coordinates. Details will be given in \cref{sec:apply-ds}.  

\subsubsection{Basic Data Structures}

\label{sec:dss}
In this section, we will review several existing constructions of classical history-independent data structures. 

Let $D$ be a classical data structure using $\tilde O(1)$ many random coins $\mathsf{r}$ that maintains a dynamically changing data set $S$. We say $D$ is \emph{history-independent} if for each possible $S$ and $\mathsf{r}$, the data structure has a unique representation $D(S,\mathsf{r})$ in the memory.
Furthermore, we say $D$ \emph{has worst-case update time $O(T)$ with high probability}, if for every possible $S$ and update operation $S\to S'$, with high probability over $\mathsf{r}$, the time complexity to update from $D(S,\mathsf{r})$ to $D(S',\mathsf{r})$ is $O(T)$. Similarly we can define worst-case query time with high probability.

Since our quantum walk algorithm is over the Johnson graph $\binom{[m]}{r}$,  for consistency we will use $r$ to denote the size of the data structure instances in the following statements.

\paragraph*{Hash tables.} We use hash tables to implement efficient lookup operations without using too much memory.
\begin{lem}[Hash tables]
\label[lemma]{lem:hashtable}
There is a history-independent data structure of size $\tilde O(r)$ that maintains a set of at most $r$ key-value pairs $\{(\mathsf{key}_1,\mathsf{value}_1),(\mathsf{key}_2,\mathsf{value}_2),\dots,(\mathsf{key}_r,\mathsf{value}_r)\}$ where $\mathsf{key}_i$'s are distinct integers from $[m]$, and supports the following operations in worst-case $\tilde O(1)$ time with high probability:
\begin{itemize}
    \item \textbf{Lookup:} Given a $\mathsf{key}\in [m]$, find the $\mathsf{value}$ corresponding to $\mathsf{key}$ (or report that $\mathsf{key}$ is not present in the set).
    \item \textbf{Insertion:} Insert a key-value pair into the set.
    \item \textbf{Deletion:} Delete a key-value pair from the set.
\end{itemize}
\end{lem}
\begin{proof}(Sketch)
  The construction is similar to \cite[Section 6.2]{DBLP:journals/siamcomp/Ambainis07}.
 The hash table has $r$ buckets, each with the capacity for storing $O(\log m)$ many key-value pairs. A pair $(\mathsf{key},\mathsf{value})$ is stored in the $h(\mathsf{key})^{\text{th}}$ bucket, and the pairs inside each bucket are sorted in increasing order of keys. 
 If some buckets overflow,  we can collect all the leftover pairs into a separate buffer of size $r$ and store them in sorted order. 
 This ensures that any set of $r$ key-value pairs has a unique representation in the memory. And, each basic operation can be implemented in $\polylog(m)$ time, unless there is an overflow.  Using an $O(\log m)$-wise independent hash function $h\colon [m] \to [r]$, for any possible $r$-subset of keys, with high probability none of the buckets overflow.\footnote{We remark that Ambainis only used a fixed hash function $h(i)= \lfloor r\cdot i/m\rfloor$, which ensures the buckets do not overflow with high probability \emph{over a random $r$-subset $K\subseteq [m]$ of keys}. Ambainis showed that this property is already sufficient for the correctness of the quantum walk algorithm.
 Here we choose to state a different version that achieves high success probability for \emph{every fixed} $r$-subset of keys, merely for keeping consistency with later presentation.}  
 \end{proof}

\paragraph*{Dynamic arrays.} We will need a dynamic array that supports indexing, insertion, deletion, and some other operations. 

The \emph{skip list} \cite{pugh} is a probabilistic data structure which is usually used as an alternative to balanced trees, and satisfies the history-independence property.
 Ambainis' quantum Element Distinctness algorithm \cite{DBLP:journals/siamcomp/Ambainis07} used the skip list to maintain a \emph{sorted} array, supporting insertions, deletions, and binary search. 
In order to apply the skip list in the quantum walk, a crucial adaptation in Ambainis' construction is to show that the random choices made by the skip list can be simulated using $O(\log n)$-wise independent functions \cite[Section 6.2]{DBLP:journals/siamcomp/Ambainis07}, which only take $\polylog(n)$ random coins to sample.
In the recent quantum fine-grained reduction result by Buhrman, Loff, Patro, and Speelman \cite[Section 3.2]{DBLP:journals/corr/abs-2106-02005}, they used a more powerful version of skip lists that supports \emph{efficient indexing}.
We will use this version of skip lists with some slight extension.

\begin{lem}[Dynamic arrays]
\label[lemma]{lem:skiplist}
There is a history-independent data structure of size $\tilde O(r)$ that maintains an array of items $(\mathsf{key}_1,\mathsf{value}_1),(\mathsf{key}_2,\mathsf{value}_2),\dots,(\mathsf{key}_r,\mathsf{value}_r)$ with distinct keys (note that neither the keys nor the values are necessarily sorted in increasing order), and supports the following operations with worst-case $\tilde O(1)$ time complexity and high success probability:
\begin{itemize}
    \item \textbf{Indexing:} Given an index $1\le i\le r $, return the $i^{\text{th}}$ item $(\mathsf{key}_i,\mathsf{value}_i)$
    \item \textbf{Insertion:} Given an index $1\le i \le r+1$ and a new item, insert it into the array between the $(i-1)^{\text{st}}$ item and the $i^{\text{th}}$ item (shifting later items to the right).
    \item \textbf{Deletion:} Given an index $1\le i\le r $, delete the $i^{\text{th}}$ item from the array (shifting later items to the left).
    \item \textbf{Location:} Given a $\mathsf{key}$, return its position $i$ in the array (i.e., $\mathsf{key}_i = \mathsf{key}$).
    \item \textbf{Range-minimum query:} Given $1\le a\le b\le r$, return $\min_{a\le i\le b}\{\mathsf{value}_i\}$.
\end{itemize}
\end{lem}
\begin{proof} (Sketch) 
We will use (a slightly modified version of) the data structure described in \cite[Section 3.2]{DBLP:journals/corr/abs-2106-02005}, which extends the construction of \cite[Section 6.2]{DBLP:journals/siamcomp/Ambainis07} to support \emph{insertion, deletion, and indexing}.
Their construction is a (bidirectional) skip list of the items, where a pointer (a ``skip'') from an item $(\mathsf{key},\mathsf{value})$ to another item  $(\mathsf{key}',\mathsf{value}')$ is stored in a hash table as a key-value  pair $(\mathsf{key},\mathsf{key}')$. To support efficient indexing, for each pointer they also store the \emph{distance} of this skip, which is used during an indexing query to keep track of the current position after following the pointers (similar ideas were also used in, e.g., \cite[Section 3.4]{pugh1998skip}). 
After every insertion or deletion, the affected distance values are updated recursively, by decomposing a level-$i$ skip into $O(\log n)$ many level-$(i-1)$ skips.  

A difference between their setting and ours is that they always keep the array sorted in increasing order of $\mathsf{value}$'s, and the position of an inserted item is decided by its relative order among the values in the array, instead of by a given position $1\le i\le r+1$.  
Nevertheless, it is straightforward to adapt their construction to our setting, by using the distance values of the skips to keep track of the current position, instead of  by comparing the values of items.

Note that using the distance values we can also efficiently implement the \emph{Location} operation in a reversed way compared to \emph{Indexing}, by following the pointers backwards and moving up levels.

To implement the \emph{range-minimum query} operations, we maintain the range-minimum value of each skip in the skip list, in a similar way to maintaining  the distance values of the skips.
They can also be updated recursively after each update. Then, to answer a query, we can  travel from the $a^{\text{th}}$ item to the $b^{\text{th}}$ by following the pointers (this is slightly trickier if $a \neq 1$, where we may first move up levels and then move down).
\end{proof}

We also need a 2D range sum data structure for points with integer coordinates.
\begin{lem}[2D range sum]
\label[lemma]{lem:range} Let integer $N \le n^{O(1)}$. There is a history-independent data structure of size $\tilde O(r)$ that maintains a multi-set of at most $r$ points $\{(x_1,y_1),\dots,(x_r,y_r)\}$ with integer coordinates $x_i\in [N], y_i \in [N]$, and supports the following operations with worst-case $\tilde O(1)$ time complexity and high success probability:
\begin{itemize}
    \item \textbf{Insertion:} Add a new point $(x,y)$ into the multiset (duplicates are allowed).
    \item \textbf{Deletion:} Delete the point $(x,y)$ from the multiset (if it appears more than once, only delete one copy of them).
    \item \textbf{Range sum:} Given $1\le x_1\le x_2\le N, 1\le y_1\le y_2\le N$, return the number of points $(x,y)$ in the multiset that are in the rectangle $[x_1\dd x_2] \times [y_1\dd y_2]$.
    \end{itemize}
\end{lem}

\begin{proof}(Sketch)
Without loss of generality, assume $N$ is a power of two. We use a simple folklore construction that resembles a 2D segment tree (sometimes called 2D range tree or 2D radix tree). Define a class $\caC = \caC_1\cup \caC_2\cup \dots \cup \caC_{\log N}$ of sub-segments of the segment $[1\dd N]$ as follows: 
\begin{align*}
    \caC_1 &= \{ [1\dd N]\},\\
    \caC_2 &= \{[1\dd N/2],[N/2+1 \dd N] \},\\
    \caC_3 &= \{[1\dd N/4],[N/4+1\dd 2N/4],[2N/4+1\dd 3N/4],[3N/4+1 \dd N]\},\\
    \dots &\\
    \caC_{\log N} &= \{[1\dd 1],[2\dd 2],\dots,[N\dd N]\}.
\end{align*}
Then it is not hard to see that every segment $[a\dd b]\subseteq [1\dd N]$ can be represented as the disjoint union of at most $2\log N$ segments in $\caC$. Consequently, the query rectangle
$[x_1\dd x_2] \times [y_1\dd y_2]$
can always be represented as the disjoint union of $O(\log^2 N)$ rectangles of the form $\mathcal{I} \times \mathcal{J}$ where $\mathcal{I}, \mathcal{J} \in \caC$. 

Hence, for every $\mathcal{I}, \mathcal{J} \in \caC$ with \emph{non-zero} range sum $s(\mathcal{I} \times \mathcal{J})$, we store this range sum into a hash table, indexed by the canonical encoding of $(\mathcal{I},\mathcal{J})$. 
Then we can efficiently answer all the range-sum queries by decomposing the rectangles and summing up their stored range sums. 

When a point $(x,y)$ is updated, we only need to update the range sums of $\log^2 N$ many rectangles that are affected, since each $a\in [1\dd N]$ is only included by $\log N$ intervals in $\caC$. We may also need to insert a new rectangle into the hash table,  or remove a rectangle once its range sum becomes zero.
\end{proof}

\paragraph*{Data structures in quantum walk.}
Ambainis \cite{DBLP:journals/siamcomp/Ambainis07} showed that a history-independent classical data structure $D$ with worst-case time complexity $T$ (with high probability over the random coins $\mathsf{r}$) can be applied to the quantum walk framework by creating a uniform superposition over all possible $\mathsf{r}$, i.e., the data structure storing data $S$ corresponds to the quantum state $\sum_{\mathsf{r}} \lvert{D(S,\mathsf{r})}\rangle \lvert{\mathsf{r}}\rangle$. During the quantum walk algorithm, each data structure operation is aborted after running for $T$ time steps. By doing this, some components in the quantum state may correspond to malfunctioning data structures, but Ambainis showed that this will not significantly affect the behavior of the quantum walk algorithm.
We do not repeat the error analysis here, but instead refer interested readers to the proof of \cite[Lemma~5~and~6]{DBLP:journals/siamcomp/Ambainis07} (see also \cite[Lemma~1~and~2]{DBLP:journals/corr/abs-2106-02005}).

\subsubsection{Applying the data structures}
\label{sec:apply-ds}
Now we will use the data structures described in \cref{sec:dss} to implement our quantum walk algorithm from \cref{sec:lcs-warm-up} time-efficiently.

Recall that $C$ is the $T$-quantum-time-constructible good anchor set of size $|C|=m$ (\cref{defn:good-anchor}). 
The states of our quantum walk algorithms are $r$-subsets $K=\{k_1,k_2,\dots,k_r\}\subseteq [m]$, where each index $k\in K$ is associated with an anchor $C(k)\in [n]$, which specifies the color (red or blue) of $k$ and the pair $(P(k),Q(k))$ of strings of lengths at most $d$.
We need to maintain the lexicographical orderings $(k_1^P,\dots,k_r^P)$ and LCP arrays $(h_1^P,\dots,h_{r-1}^P)$, so that $P(k_1^P) \preceq P(k_2^P)\preceq \cdots  \preceq P(k_r^P)$ and $h_i^P = \lcp(P(k_i^P),P(k_{i+1}^P))$, and similarly maintain $(k_1^{Q},\dots,k_r^{Q}),(h_1^Q,\dots,h_{r-1}^Q)$ for the strings $\{Q(k)\}_{k\in K}$.

For $k\in K$, we use $\pos^P(k)$ to denote the position $i$ such that $k^P_i = k$, i.e., the lexicographical rank of $P(k)$ among all $P(k_1),\dots,P(k_r)$. Similarly, let $\pos^Q(k)$ denote the position $i$ such that $k^{Q}_i = k$.

We can immediately see that all the steps in the \emph{update step} (Algorithm \ref{algo:insert}) of our quantum walk can be implemented time-efficiently. In particular, we use a hash table (\cref{lem:hashtable}) to store the anchor $C(k)$ corresponding to each $k\in K$, and use \cref{lem:skiplist} to maintain the lexicographical orderings and LCP arrays under insertions and deletions. Each update operation on these data structures takes $\tilde O(1)$ time. 
Additionally, these data structures allow us to efficiently compute some useful information, as summarized below.
\begin{prop}
\label[proposition]{prop:cancompute}
Given indices $k,k' \in K$, the following information can be computed in $\tilde O(1)$ time.
\begin{enumerate}
    \item The anchor $C(k)$, the color of $k$, and the lengths  $|P(k)|,|Q(k)|\le d$.\label{item:1}
    \item $\pos^P(k)$ and $\pos^Q(k)$.\label{item:2}
    \item $\lcp(P(k),P(k'))$ and $\lcp(Q(k),Q(k'))$.\label{item:3}
\end{enumerate}
\end{prop}
\begin{proof}
For \cref{item:1}, rather than use $T$ time to compute $C(k)$ (\cref{defn:good-anchor}), we instead look up the value of $C(k)$ from the hash table. Then, $C(k)\in[n]$ determines the color of $k$ and the string lengths.

For \cref{item:2}, we use the location operation of the dynamic array data structure (\cref{lem:skiplist}).

For \cref{item:3}, we first compute $i = \pos^P(k), i' = \pos^P(k')$, and  assume $i<i'$ without loss of generality. 
Then, by \cref{lem:lcp-height}, we can compute $\lcp(P(k),P(k')) = \lcp(P(k^P_i),P(k^P_{i'})) = \min\{h^P_{i},h^P_{i+1},\dots,h^P_{i'-1}\}$ using a range-minimum query (\cref{lem:skiplist}).
\end{proof}

\newcommand{\kred}{k^{\mathsf{red}}}
\newcommand{\kblue}{k^{\mathsf{blue}}}
The remaining task is to efficiently implement the checking step, where we need to solve the Two String Families LCP problem. The goal is to find a red index $\kred \in K$ and a blue index $\kblue \in K$, such that $\lcp(P(\kred),P(\kblue))+ \lcp(Q(\kred),Q(\kblue))\ge d$. Now we give an outline of the algorithm for solving this task.

\begin{algorithm2e}
\label{algo:checking}
\DontPrintSemicolon
  \SetKwFor{Search}{Grover-Search}{}{}
	\Search{over red indices $\kred \in K$, and integers $d'\in [0\dd d] $} {
	Find $\ell^P, r^P$ such that $\lcp(P(k_{i}^P), P(k^{\mathsf{red}}))\ge d'$ if and only if $\ell^P\le i\le r^P$. \label{line:findplr}\\
	Find $\ell^Q, r^Q$ such that $\lcp(Q(k_{i}^{Q}), Q(k^{\mathsf{red}}))\ge d-d'$ if and only if $\ell^Q\le i\le r^Q$. \label{line:findqlr}\\
	\lIf{exists a blue index $\kblue \in K$ such that $\pos^P(k^{\mathsf{blue}}) \in [\ell^P\dd r^P], \pos^Q(k^{\mathsf{blue}}) \in [\ell^Q\dd r^Q]$ \label{line:check}}  {
	    \Return{True}
	}
	}
	\Return{False}
	\caption{Solving the Two String Families LCP problem in the checking step}
\end{algorithm2e}

In the Algorithm~\ref{algo:checking}, we use Grover search to find a red index $\kred \in K$ and an integer $d'\in [0\dd d]$, such that there exists a blue index $\kblue \in K$ with $\lcp(P(\kred),P(\kblue))\ge d'$ and $\lcp(Q(\kred),Q(\kblue))\ge d-d'$. The number of Grover iterations is $\tilde O(\sqrt{|K|\cdot d}) = \tilde O(\sqrt{rd})$, and we will implement each iteration in $\polylog(n)$ time. 
By \cref{lem:lcp-height}, all the strings $P(k)$ that satisfy $\lcp(P(k),P(\kred))\ge d'$ form a \emph{contiguous segment} in the lexicographical ordering $P(k_1^P)\preceq\dots\preceq P(k_r^P)$. 
In Line~\ref{line:findplr}, we find the left and right boundaries $\ell^P, r^P$ of this segment, using a binary search with \cref{prop:cancompute} (\cref{item:3}).  Line~\ref{line:findqlr} is similar to Line~\ref{line:findplr}. Then, Line~\ref{line:check} checks the existence of such a blue string pair.
It is clear that this procedure correctly solves the Two String Families LCP problem. The only remaining problem is how to implement Line~\ref{line:check} efficiently.

Note that Line~\ref{line:check} can be viewed as a 2D orthogonal range query, where each 2D point is a blue string pair $(P(k),Q(k))$, with coordinates being strings to be compared in lexicographical order. We cannot simply replace the coordinates by their ranks $\pos^P(k)$ and $\pos^Q(k)$ among the $r$ substrings in the current state, since their ranks will change over time. It is also unrealistic to replace the coordinates by their ranks among all the possible substrings $\{P(k)\}_{k\in [m]}$, since $m$ could be much larger than the desired overall time complexity $n^{2/3}$.  These issues seem to require our 2D range query data structure to be comparison-based, which is also difficult to achieve as mentioned before.

Instead, we will use a sampling technique, which effectively converts the non-integer coordinates into integer coordinates.  
At the very beginning of the algorithm (before running the quantum walk), we uniformly sample $r$ distinct indices $x_1,x_2\dots,x_r\in [m]$, and sort them so that $P(x_1)\preceq P(x_2) \preceq\cdots \preceq P(x_r)$ (breaking ties by the indices), in $\tilde O(r(\sqrt{d}+T))$ total time (this complexity is absorbed by the time complexity of the setup step $\sfS = O(r(\sqrt{d}+T))$). 
 Then, during the quantum walk algorithm, when we insert an index $k\in [m]$ into $K$, we assign it an \emph{integer label} $\rho^P(k)$ defined as the unique $i \in [0\dd r]$ satisfying $P(x_i)\preceq s' \prec P(x_{i+1})$, which can be computed in $\tilde O(\sqrt{d})$ time by a binary search on the sorted sequence $P(x_1)\preceq \cdots \preceq P(x_r)$.
 We also sample $y_1,\dots,y_r\in [m]$ and sort them so that $Q(y_1)\preceq Q(y_2)\preceq \dots \preceq Q(y_r)$, and similarly define the integer labels $\rho^{Q}(k)$.
 Intuitively, the (scaled) label $\rho^P(k) \cdot (m/r)$ estimates the rank of $P(k)$ among all the strings $\{P(k')\}_{k'\in [m]}$.
 
The following lemma formalizes this intuition. It states that in a typical $r$-subset $K = \{k_1,k_2,\dots,k_r\}\subseteq [m]$, not too many indices can receive the same label.
\begin{lem}
\label[lemma]{lem:balls}
For any $c>1$, there is a $c'>1$, such that the following statement holds:

For positive integers $r\le m$,  let $A,B\subseteq [m]$ be two independently uniformly random $r$-subsets. 
	Let $A = \{a_1,a_2,\dots,a_r\}$ where $a_1<a_2<\dots<a_r$, and denote \[A_0:=[1\dd a_1), A_1:=[a_1\dd a_2),\dots,A_{r-1}:=[a_{r-1}\dd a_r), A_{r}:=[a_r\dd m].\]
	Then, 
	\[  \Pr_{A,B}\big [\text{$|A_i\cap B|\ge c'\log m \ $ for some $0\le i\le r$}\big ] \le \frac{1}{m^{c}}.\]
\end{lem}
\begin{proof}
	Let $k = c' \log m$ for some $c'>1$ to be determined later, and we can assume $k\le r$. Observe that, $|A_i\cap B| \ge k$ holds for some $i$ only if there exist $b,b'\in [m]$, such that $|[b\dd b'] \cap B| \ge k$ and $[b+1\dd b'] \cap A = \emptyset$. 
	
	Let $b,b'\in [m], b\le b'$. For $b'-b \ge (c+2) (m\ln m)/r$, we have
	\begin{align*}
	    \Pr_A[  [b+1\dd b']\cap A = \emptyset] &= \frac{\binom{m-(b'-b)}{r}}{\binom{m}{r}}\\
	    &\le \left (1 -  \frac{b'-b}{m}\right )^r\\
	    & \le 1/m^{c+2}.
	\end{align*} 
	For $b'-b < (c+2)(m\ln m)/r$,   we have
	\begin{align*}
	\Pr_B\big [| [b\dd b']\cap B |\ge k\big ]  & \le \frac{\binom{b'-b+1}{k} \cdot \binom{m-k}{r-k}}{\binom{m}{r}}\\ &= \binom{b'-b+1}{k} \cdot \frac{\binom{r}{k}}{\binom{m}{k}}\\ &\le \left (\frac{e(b'-b+1)}{k}\right )^k \cdot \left (\frac{r}{m}\right )^k\\
	& < \left ( \frac{e (c+3) \ln m }{k}\right )^k\\
	&< 1/m^{c+3},
	\end{align*}
	where we set $k = c'\log m = 3(c+3)\log m$.

The proof then follows from a union bound over all pairs of $b,b'\in [m]$.
\end{proof}

Then, we can use the 2D point $(\rho^P(k),\rho^Q(k))$ with integer coordinates to represent the string pair $(P(k),Q(k))$, and use the data structure from \Cref{lem:range} to handle the 2D range sum queries. To correctly handle the points near the boundary of a query, we need to check them one by one, and \cref{lem:balls} implies that in average case this brute force step is not expensive. 

The pseudocode in Algorithm~\ref{algo:extra-insert} describes the additional steps to be performed during each insertion step of the quantum walk (the deletion step is simply the reversed operation of the insertion step).
\begin{algorithm2e}
\label{algo:extra-insert}
	Given an index $k\in [m]$ \\		Compute the integer labels $\rho^P(k)$ and $\rho^Q(k)$ using binary search, and store them in hash table\\ 
	\If{$k$ is  blue}{
	Insert the 2D point $(\rho^P(k),\rho^Q(k))$ into the 2D range sum data structure
	}
	\caption{Extra steps in the insertion procedure (in addition to the steps in  Algorithm~\ref{algo:insert})}
\end{algorithm2e}

The pseudocode in Algorithm~\ref{algo:line4} describes how to implement Line~\ref{line:check} in Algorithm~\ref{algo:checking} for solving the Two String Families LCP problem.
Line~\ref{line:internal} correctly handles all the ``internal'' blue pairs $(P(\kblue),Q(\kblue))$, which must satisfy $\pos^P(\kblue) \in [\ell^P,r^P]$ and $\pos^Q(\kblue) \in [\ell^Q,r^Q]$ by the definition of our integer labels $\rho^P(\cdot),\rho^Q(\cdot)$ and Lines~\ref{line:1p} and \ref{line:1q}.
In Line~\ref{line:forloop} we handle the remaining possible blue pairs, which must have $\rho^P(\kblue) \in \{\tilde \ell^P,\tilde r^P\}$ or $\rho^Q(\kblue) \in \{\tilde \ell^Q,\tilde r^Q\}$, and can be found by binary searches on the lexicographical orderings (to be able to do this, we need to maintain the lexicographical orderings of $P(k_1),\dots,P(k_r)$ and the sampled strings $P(x_1),\dots,P(x_r)$ \emph{combined}).

\begin{algorithm2e}
\DontPrintSemicolon
\label{algo:line4}
	Given indices $\ell^P,r^P,\ell^Q,r^Q$. \\			
	Let $\tilde \ell^P := \rho^P(k^P_{\ell^P}),\tilde r^P := \rho^P(k^P_{r^P})$.\label{line:1p}\\ 
	Let $\tilde \ell^Q := \rho^Q(k^Q_{\ell^Q}),\tilde r^Q := \rho^Q(k^Q_{r^Q})$.\label{line:1q}\\ 
	\lIf{the 2D range sum of $[\tilde \ell^P+1\dd \tilde r^P-1]\times [\tilde \ell^Q+1\dd \tilde r^Q-1]$ is non-zero}{\Return{True} \label{line:internal}}
	\For{blue index $\kblue \in K$ such that $\rho^P(\kblue) \in \{ \tilde \ell^P, \tilde r^P\}$ or $\rho^Q(\kblue) \in \{ \tilde \ell^Q, \tilde r^Q\}$\label{line:forloop}}{
	\lIf {$\pos^P(k^{\mathsf{blue}}) \in [\ell^P\dd r^P], \pos^Q(k^{\mathsf{blue}}) \in [\ell^Q\dd r^Q]$}{\Return{True}}  
	\lIf{already looped $4 c'\log m$ times}{exit for loop \label{line:abort}} 
	}
	\Return{False}
	\caption{Implementation of Line~\ref{line:check} in  Algorithm~\ref{algo:checking}}
\end{algorithm2e}

Note that in Line~\ref{line:abort} of Algorithm~\ref{algo:line4} we abort if we have checked more than $4c'\log m$ boundary points, so that Algorithm~\ref{algo:line4} has worst-case $\tilde O(1)$ overall running time.
But this early stopping would also introduce (one-sided) error if there are too many boundary points which we have no time to check.
However, a straightforward application of \cref{lem:balls} implies that, with high success probability over the initial samples $P(x_1)\preceq P(x_2)\preceq \cdots \preceq P(x_r)$ and $Q(y_1)\preceq Q(y_2)\preceq \cdots \preceq Q(y_r)$, only $1/\poly(m)$ fraction of the $r$-subsets $K= \{k_1,\dots,k_r\} \in [m]$  in the Johnson graph can have more than $c'\log m$ strings receiving the same label. 
On these problematic states
$K=\{k_1,\dots,k_r\} \in [m]$
, the checking procedure may erroneously recognize $K$ as unmarked, while other states are handled correctly by Algorithm~\ref{algo:line4} since there is no early aborting. This decreases the fraction of marked states in the Johnson graph by  only a $1/\poly(m)$ fraction, which does not affect the overall time complexity of our quantum walk algorithm.

\subsection{Improved Construction of Good Anchor Sets}
\label{sec:anchor}

In this section, we will prove \Cref{lem:good-anchor-} by constructing a good anchor set with smaller size.
Our construction of good anchor sets is based on a careful combination of a generalized version of \emph{difference covers} \cite{DBLP:conf/cpm/BurkhardtK03,DBLP:journals/tocs/Maekawa85} and the \emph{string synchronizing sets} \cite{bwt}.

\subsubsection{Approximate Difference Covers}
\label{sec:diffcover}
We first need to generalize the notion of difference covers. 

\begin{definition}[Approximate Difference Covers]
\label{defn:dlcover}
A set $D\subseteq \N^+$ is called a $(d,\tau)$-cover, if for every $i,j\in \N^+$, there exists two integers $h_1(i,j),h_2(i,j) \in [0\dd d)$ such that $i+h_1(i,j),j+h_2(i,j) \in D$, and $|h_1(i,j)-h_2(i,j)| \le \tau -1$.
\end{definition}
The notion of $d$-cover (\Cref{defn:dlcover-easy}) used in previous algorithms corresponds to the $\tau=1$ case of our new definition. 
Our generalization to larger $\tau$ can be viewed as an approximate version of difference covers with additive error $\le \tau-1$. As we shall see, allowing additive error makes the size of the $(d,\tau)$-cover much smaller compared to \Cref{defn:dlcover-easy}. 

We present a construction of approximate difference covers, by adapting previous constructions from $\tau=1$ to general values of $\tau$. 
\begin{lem}[Construction of $(d,\tau)$-cover]
\label[lemma]{lem:construct-dcover}
For every positive integers $1\le \tau \le  d$, there is a $(d,\tau)$-cover $D$ such that $D\cap [n]$ contains $O(n/\sqrt{d\tau })$ elements. 
Moreover, given integer $i\ge 1$, one can compute the $i^{\text{th}}$ smallest element of $D\cap [n]$ in $\tilde O (1)$ time.
\end{lem}
\begin{proof}
Let $M:= \left \lfloor \sqrt{d/\tau} \right \rfloor \ge 1$.
Define \[I: = \{ z M \cdot \tau \mid z\in \N^+\}, \] and  \[J := \{(xM^2-y)\cdot \tau \mid  x\in \N^+, y\in [M]\}.\]
We claim that $D:= I\cup J$ is a $(d,\tau)$-cover that satisfies the desired properties.

First, observe that $|I\cap [n]|=\lfloor  n/(M\tau)\rfloor \le O(n/\sqrt{d \tau})$, and $|J\cap [n]|\le \lfloor n/\tau\rfloor \cdot (M/M^2) = O(n/\sqrt{d \tau})$. Hence $D=I\cup J$ satisfies the claimed size bound.

Next, we verify $D$ is indeed a $(d,\tau)$-cover. For any $i,j\in \N^+$, let $i':= \lceil i/\tau\rceil, j':= \lceil j/\tau\rceil$. Let $z\cdot\tau \in J$ be the smallest integer in $J$ such that $z\ge j'$ and $z\equiv j'-i' \pmod{M}$. By the construction of $J$, we have $z\le j'+M^2-1$. Hence, let $h_1(i,j)=(z-j'+i')\cdot \tau - i$ and $h_2(i,j) = z\cdot \tau - j$. Note that 
\begin{align*}
    h_2(i,j) &\le (j'+M^2-1)\cdot \tau - j \le M^2\tau -1 \le d-1,\\
    h_2(i,j) &\ge j'\cdot \tau - j \ge 0,
\end{align*}
where we used $j' \tau - j \in [0\dd \tau-1 ]$. Similarly we can show $0\le h_1(i,j) \le d-1$, and we have 
\[
    |h_2(i,j)-h_1(i,j)| = |(j' \tau - j) - (i' \tau - i)|\le \tau -1.
    \]
Moreover, $j+h_2(i,j)\in J\subseteq D$, and \[i+h_1(i,j) = (z-j'+i')\cdot \tau \equiv 0 \pmod{M\tau}\] which implies $i+h_1(i,j)\in I\subseteq D$.  
\end{proof}

\subsubsection{String Synchronizing}
\label{sec:sync}
In \Cref{cor:good-anchor-easy} we obtained a good anchor set using a $(d,1)$-cover. If we simply replace it by a $(d,\tau)$-cover with larger $\tau$, the size of the obtained anchor set would become smaller, but it would no longer be a good anchor set, due to the misalignment introduced by approximate difference covers. To deal with the misalignment, we will use the \emph{string synchronizing sets} recently introduced by Kempa and Kociumaka \cite{bwt}. Informally, a synchronizing set of string $S$ is a small set of synchronizing positions, such that every two sufficiently long matching substrings of $S$ with no short periods should contain a pair of consistent synchronizing positions.

\begin{definition}[String synchronizing sets  {\cite[Definition 3.1]{bwt}}]
For a string $S[1\dd n]$ and a positive integer $1\le \tau \le n/2$, we say $A\subseteq [1\dd n-2\tau+1]$ is a \emph{$\tau$-synchronizing set of $S$} if it satisfies the following properties:
\begin{itemize}
    \item \textbf{Consistency:} If $S[i\dd i+2\tau) = S[j\dd j+2\tau)$, then $i\in A$ if and only if $j\in A$. 
    \item \textbf{Density:} For $i\in [1\dd n-3\tau +2]$, $A \cap [i\dd i+\tau) =\emptyset$ if and only if $\per(S[i\dd i+3\tau - 2])\le \tau/3$.
\end{itemize}
\end{definition}

\begin{comment}
For a $\tau$-synchronizing set $A$ of $S[1\dd n]$, denote $\suc_{A}(i) := \min \{j \in A\cup \{n-2\tau+2\} \mid j\ge i\}$. 
Then we have the following lemmas proved in \cite{bwt}.
\begin{lem}[{\cite[Fact 3.2]{bwt}}]
If $p = \per(S[i\dd i+3\tau-2])\le \tau/3$, then $S[i\dd \suc_A(i)+2\tau -1 )$ is the longest prefix of $S[i\dd n]$ with period $p$.
\end{lem}
\begin{lem}[{\cite[Fact 3.3]{bwt}}]
If a string $U$ with $|U|\ge 3\tau -1$ and $\per(U)>\tau/3$ occurs at positions $i$ and $j$ in $S$, then $\suc_{A}(i)-i = \suc_{A}(j)-j \le |U| - 2\tau$.
\end{lem}
\end{comment}

Kempa and Kociumaka gave a linear-time classical randomized algorithm (as well as a derandomized version, which we will not use here) to construct a $\tau$-synchronizing set $A$ of optimal size\footnote{In the case where $S$ has no highly periodic substrings, every $\tau$-length interval should contain at least one index from $A$.} $|A|=O(n/\tau)$. 
However, this classical algorithm for constructing $A$
has to query each of the $n$ input characters, and is not directly applicable to our sublinear quantum algorithm.

To apply Kempa and Kociumaka's construction algorithm to the quantum setting, we observe that this algorithm is \emph{local}, in the sense that whether an index $i$ should be included in $A$ is completely decided by its short context $S[i\dd i+2\tau)$ and the random coins. Moreover, by suitable adaptation of their construction, one can compute all the synchronizing positions in an $O(\tau)$-length interval in $\tilde O(\tau)$ time. 
We summarize all the desired properties of the synchronizing set in the following lemma.
\begin{lem}[Adaptation of \cite{bwt}]
\label[lemma]{lem:kk-property}
For a string $S[1\dd n]$ and a positive integer $1\le \tau \le n/2$, given a sequence $\sf{r}$ of $O(\log n)$ many random coins, there exists a set $A$ with the following properties:
\begin{itemize}
    \item \textbf{Correctness:} With high probability over $\sf{r}$, $A$ is a $\tau$-synchronizing set of $S$.
    \item \textbf{Locality:}  For every $i\in [1\dd n-2\tau +1]$, whether $i\in A$ or not is completely determined by the random coins $\sf{r}$ and the substring $s[i\dd i+2\tau)$. 
    
    Moreover, given $s[i\dd i+4\tau)$ and $\sf{r}$, one can compute all the elements in $A\cap [i\dd i+2\tau)$ by a classical algorithm in $\tilde O(\tau)$ time.
    \item \textbf{Sparsity:}  
    For every $i\in [1\dd n-2\tau+1]$,   $\Ex_{\sf {r}}\big [|A\cap [i\dd i+2\tau)|\big ] \le 80$.
\end{itemize}
\end{lem}

In the following, we first inspect  the (slightly adapted) randomized construction of string synchronized sets by Kempa and Kociumaka \cite{bwt}, and then show that it satisfies the properties in \Cref{lem:kk-property}. 
\paragraph*{Construction of string synchronizing sets.} Fix an input string $S[1\dd n]$ and a positive integer $\tau \le n/2$.
Define sets
\begin{align*}
Q &= \{i\in [1\dd n-\tau+1]: \per (S[i\dd i+\tau))\le \tau/3 \},\\
 B &= \big \{i\in [1\dd n-\tau+1]\setminus Q: \per (S[i\dd i+\tau-1))\le \tau/3\, \lor\, \per (S[i+1\dd i+\tau) )\le \tau/3\big \},
\end{align*}
where we define $B=\emptyset$ in the special case of $\tau=1$.

Let $\caP = \{s\in \Sigma^{\tau}: \text{$s$ is a substring of $S$}\}$ denote the set of all the length-$\tau$ substrings in $S$ (without duplicates). Let $\pi\colon \caP \to [N]$ be any injection, and define the identifier function $\id\colon [1\dd n-\tau+1]\to \N^+$ by
\[\id(i):= \begin{cases}\pi\big (S[i\dd i+\tau)\big ) & i\in B, \\ \pi\big (S[i\dd i+\tau)\big ) + N & i \notin B. \end{cases}\]
In this way, we have $\id(i)=\id(j)$ if and only if $S[i\dd i+\tau) = S[j\dd j+\tau)$. Moreover, for $i\in B,j\notin B$, we always have $\id(i)<\id(j)$. Finally, define
 \[ A = \big \{i \in [1\dd n-2\tau+1]:\min\{\id(j): j\in [i\dd i+\tau]\setminus Q\} \in \{\id(i),\id(i+\tau)\}\big \}.\]
Kempa and Kociumaka proved the following fact.
\begin{lem}[{\cite[Lemma 8.2]{bwt}}]
\label[lemma]{lem:correctnes}
The set $A$ is always a $\tau$-synchronizing set of string $S$.
\end{lem}

We first quickly verify the locality property of this construction.
\begin{prop}
\label[proposition]{prop:quick-locality}
For every $i\in [1\dd n-2\tau +1]$, whether $i\in A$ or not is completely determined by $\pi$ and the substring $s[i\dd i+2\tau)$.
\end{prop}
\begin{proof}
    This immediately follows from the definition of $Q,B,\id,$ and $A$. 
\end{proof}

Now, suppose $\pi\colon \caP \to [N]$ is randomly chosen so that the \emph{$0.1$-approximate  min-wise independence} property is satisfied: for any $x\in \caP$ and subset $X\subseteq \caP \setminus \{x\}$, 
\[ \Pr_{\pi} \big [\pi(x) < \min \{\pi(x'): x' \in X\}\big ] \in \frac{1}{|X|+1}\cdot (1\pm 0.1).\]
Then the following holds.
\begin{lem}[{\cite[Fact 8.9]{bwt}}, adapted] The expected size of $A$ satisfies 
$\Ex_{\pi}[|A|] \le 20n/\tau$.
\label[lemma]{lem:expect}
\end{lem}

\begin{remark}
We remark that in the original construction of \cite{bwt}, $\pi$ was chosen to be a uniformly random bijection $\caP\to [|\caP|]$, and this is \emph{the only part that differs from our modified version}. The main issue with this ideal choice is that, in our quantum algorithm, we do not have enough time to sample and store $\pi$, which could have size $\Omega(n)$. Observe that in their proof of \cref{lem:expect}, the only required property of $\pi$ is that $\pi$  satisfies (perfect) min-wise independence. Hence, here we can relax it to have approximate min-wise independence, and their proof of \cref{lem:expect} still applies (with a slighly worse constant factor).
\end{remark}

Now we describe how to design such a mapping $\pi$ that is efficiently computable.  First, we hash the substrings into integers using the standard rolling hash method \cite{DBLP:journals/ibmrd/KarpR87}. Recall that the alphabet $\Sigma$ is identified with the integer set $[|\Sigma|]$.

\begin{definition} [Rolling hash]
   Let $p>|\Sigma|$ be a prime, and pick $y\in \F_p$ uniformly at random. Then, the rolling hash function $\rho_{p,y}\colon \Sigma^\tau \to \F_p $ on length-$\tau$ strings is defined as
   \[ \rho_{p,x}\big ( s[1\dd \tau]\big ) := \sum_{i=1}^{\tau} s[i]\cdot y^{i} \pmod{p}.\]
\end{definition}
We have two the following two folklore facts about rolling hash.
\begin{itemize}
    \item  Rolling hashes of substrings can be easily computed in batch: on any given string $s$ of length $|s|\ge \tau$, one can compute the hash values $\rho_{p,y}\big ( s[i\dd i+\tau)\big )$ for \emph{all} $i \in [1\dd |s|-\tau+1]$, in $O(|s| \cdot \polylog p)$ total time.
 \item 
    By choosing $p = \poly(n)$, we can ensure that with high probability over the choice of $y$, the rolling hash function $\rho_{p,y}$ takes distinct values over all the strings in $\caP$ (by Schwartz-Zippel lemma).
\end{itemize}

After hashing the strings in $\caP$ to a small integer set $[\poly(n)]$, we can apply known constructions of  approximate min-wise independent hash families.
\begin{lem}[Approximate min-wise independent hash family, e.g., \cite{DBLP:journals/jal/Indyk01}]
Given parameter $n \ge 1$, one can choose $N \le  n^{O(1)}$, so that there is a hash family $\mathcal{H} = \{h\colon [N]\to [N]\}$ that satisfies the following properties:
\begin{itemize}
    \item \textbf{Injectivity:} For any subset $X\subseteq [N]$ of size $|X|\le n$, with high probability over the choice of $h \in \mathcal{H}$, $h$ maps $X$ to distinct elements.
    \item \textbf{Approximate min-wise independence:} For any $x\in [N]$ and subset $X\subseteq [N] \setminus \{x\}$, 
\[ \Pr_{h\in \mathcal{H}} \big [h(x) < \min \{h(x'): x' \in X\}\big ] \in \frac{1}{|X|+1}\cdot (1\pm 0.1).\]
    \item \textbf{Explicitness:} Each hash function $h\in \mathcal{H}$ can be specified using $O(\log n)$ bits, and can be evaluated at any point in $\polylog(n)$ time.
\end{itemize}
\end{lem}
Finally, we choose parameters $p=\poly(n), N = \poly(n),p\le N$, and define the pseudorandom mapping $\pi\colon \caP \to [N]$ by $\pi (s):= h\big (\rho_{p,y}(s)\big )$, where $\rho_{p,y}\colon \caP \to \F_p$ is the rolling hash function (identifying $\F_p$ with $[p] \subseteq [N]$), and $h\colon [N] \to [N]$ is the approximate min-wise independent hash function. 

Now we are ready to prove that the string synchronizing set $A$ determined by the random mapping $\pi$ satisfies the properties stated in  \Cref{lem:kk-property}.
\begin{proof} (of \Cref{lem:kk-property})
First note that the random coins $\sf{r}$ are used to sample $y\in \F_p$ and $h\in \mathcal{H}$, which only take $O(\log n)$ bits of seed.

\paragraph*{Correctness:} By \Cref{lem:correctnes}, $A$ is correct as long as $\pi$ is an injection, which holds with high probability by the injectivity properties of $\rho_{p,y}$ and $h$.
     \paragraph*{Locality:} The first part of the statement is already verified in \cref{prop:quick-locality}. To show the moreover part, first note that the values of $\pi\big( S[j\dd j+\tau)\big )$ over all $j\in [i\dd i+3\tau)$ can be computed in $\tilde O(\tau)$ time, by the property of rolling hash and the explicitness of $h$. 
    By \cite[Lemma 8.8]{bwt}, the sets $Q\cap [i \dd i+3\tau)$ and $B\cap [i\dd i+3\tau)$ can be computed in $O(\tau)$ time. Hence, we can compute $\id(j)$ for all $j\in [i\dd i+3\tau)$, and then construct $A\cap [i\dd i+2\tau)$ by computing the sliding-window minima, in $\tilde O(\tau)$ overall time.
     \paragraph*{Sparsity:}
Let $S'=S[i\dd i+4\tau)$, and construct an $\tau$-synchronizing set $A'$ of $S'$ using the \emph{same} random coins $\sf{r}$. Then, from the locality property we clearly have $|A'|\ge |A\cap [i\dd i+2\tau)|$.  Hence, by \Cref{lem:expect},  $\Ex_{\sf{r}}\big [|A \cap [i\dd i+2\tau)|\big ] \le \Ex_{\sf{r}}\big [|A'|] \le 20\cdot 4\tau/\tau = 80$. 
\end{proof}

\subsubsection{Putting it together}
Now we will construct the good anchor set for input strings $s,t$ and threshold length $d$.  
 Recall that $S=s\$t$ and $|S|=n$,  and we have assumed $d\ge 100$ in order to avoid corner cases.
 Our plan is to use string synchronizing sets to fix the misalignment introduced by the approximate different covers. However, in highly-periodic parts where synchronizing fails, we have to rely on periodicity arguments and Grover search.

\begin{construction}[Anchor set $C$]
Let $D$ be a $(\lfloor d/2 \rfloor,\tau)$-cover for some parameter $\tau \le d/100$ to be determined later, and let $D_{S}:=\big (D\cap [|s|]\big ) \cup \big (|s|+1+(D\cap [|t|])\big ) $.
 Let $A$ be the $\tau$-synchronizing set of $S$ determined by random coins $\sf{r}$.

For every $i\in D_S \cap [n-3\tau+2]$, let $L_i\subseteq [1\dd n]$ be defined by the following procedure.
\label[definition]{defn:anchor}
\begin{itemize}
	\item \textbf{Step 1:} If  $A \cap [i\dd i+2\tau)$ has at most 1000 elements, then add all the elements from $A \cap [i\dd i+2\tau)$ into $L_i$. Otherwise,  add the smallest 1000 elements from $A \cap [i\dd i+2\tau)$ into $L_i$.
	\item \textbf{Step 2:} If $p:=\per(S[i+\tau\dd i+3\tau-2]) \le \tau/3$, then we do the following: 
	\begin{itemize}
	    \item Define two boundary indices 
	    \begin{align*}
	        r &:= \max\big \{r: r\le \min \{n,i+d\} \land \per(S[i+\tau\dd r])=p\big \},\\ \ell &:= \min\big \{\ell: \ell\ge \min \{1,i-d\} \land \per(S[\ell\dd i+3\tau-2])=p\big \}.
	    \end{align*} Let $P$ be the Lyndon root of
	    $S[i+\tau\dd i+3\tau-2]$ (see \cref{sec:prelim:notation}). Then $|P|=p$, and let $P=S[i^{(b)}\dd i^{(b)}+p)=S[i^{(e)}\dd i^{(e)}+p)$ be the first and last occurrences of $P$ in $S[\ell \dd r]$. We add three elements $i^{(b)},i^{(b)}+p$, and $i^{(e)}$ into $L_i$.
	\end{itemize}
\end{itemize}
Finally, the anchor set $C$ is defined as $\bigcup_{i\in D_S\cap [n-3\tau+2]} L_i$.
\end{construction} 
Before proving the correctness of the anchor set $C$ in \cref{defn:anchor}, we first observe that $C$ has small size and is efficiently constructible. 

\begin{lem}
\label[lemma]{prop:local-access}
The anchor set $C$ has size $|C|\le O(n/\sqrt{d\tau})$, and is $\tilde O(\tau + \sqrt{d})$-quantum-time constructible.
\end{lem}
\begin{proof}
	For any given $i\in D_S\cap [n-3\tau+2]$, $L_i$ contains at most 1003 elements. Hence, $|C|\le 1003\cdot |D_S| \le O(n/\sqrt{d\tau})$ by \Cref{lem:construct-dcover}.
	
	In \Cref{defn:anchor}, Step 1 takes $\tilde O(\tau)$ classical time by the Locality property in \Cref{lem:kk-property}. 
	In Step 2, we can find the period $p$ and the Lyndon root $P$ in $\tilde O(\tau)$ classical time (see \cref{sec:prelim:notation}). Then, finding the right boundary $r$ is equivalent to searching for the largest $r \in \big [i+3\tau -2\dd \min \{n,i+d\}\big ]$ such that $p$ is a period of $S[i+\tau \dd r]$ (this is because we already know $\per(S[i+\tau \dd i+3\tau-2])=p$, and the period is monotonically non-decreasing in $r$). 
	We do this with a binary search over $r$, where each check can be performed by a Grover search in $\tilde O(\sqrt{d})$ time, since the length of $S[i+\tau \dd r]$ is at most $d$.  The left boundary $\ell$ can be found similarly.  Finally, the positions $i^{(b)}$ and $i^{(e)}$ can be found in $\tilde O(p)$ time classically, since we must have $i^{(b)}-\ell, r-i^{(e)} \le 2p$. Hence, our anchor set $C$ is $\tilde O(\tau + \sqrt{d})$-quantum-time constructible.
\end{proof}

Now we show that, with constant probability, $C$ is a good anchor set (\cref{defn:good-anchor}).
\begin{lem}
\label[lemma]{lem:good-anchor}
For input strings $s,t$ and threshold length $d$, with at least $0.8$ probability over the random coins $\sf{r}$, the set $C$ is a good anchor set.
\end{lem}
The proof of this lemma has a similar structure to the proof of \cite[Lemma 17]{lcs2021}, but is additionally complicated by the fact that (1)~we have to deal with the misalignment introduced by approximate difference covers, and (2)~we only considered a subset of the synchronizing set when defining the anchors. 

Here, we first give an informal overview of the proof. By the property of approximate difference covers, the length-$d$ common substring of $s$ and $t$ should have a pair of slightly misaligned anchors within a shift of  at most $\tau-1$ from each other. If the context around these misaligned anchors are not highly-periodic (Case 1 in the proof below), then their $O(\tau)$-neighborhood must contain a pair of synchronizing positions (by the density property of $A$), which are included in Step 1 of \Cref{defn:anchor}, and form a pair of perfectly aligned anchors (by the consistency property of $A$).  If the context around the misaligned anchors are highly-periodic (Case 2), we can extend the period to the left or to the right, and look at the first position where the period stops. If this stopping position is inside the common substring, then we have a pair of anchors (Cases 2(i), 2(ii)). Otherwise, the whole common substring is highly-periodic, and we can also obtain anchors by looking at the starting positions of its Lyndon roots (Case 2(iii)). These anchors for Case 2 are included in Step 2 of \Cref{defn:anchor}.
  
\begin{proof} (of \Cref{lem:good-anchor})
Let $s[i_{\star}\dd i_{\star}+d)=t[j_{\star}\dd j_{\star}+d)$ be a length-$d$ common substring of $s$ and $t$.  Our goal is to show the existence of positions $i \in [|s|-d+1],j\in [|t|-d+1]$ and a shift $h\in [0\dd d)$, such that $s[i\dd i+d)=t[j\dd j+d)$, and $i+h,  |s|+1+j+h \in C$.

 Recall that we assumed $d\ge 100\tau$.  By the definition of $D_S$, there exist $h_1, h_2\in [0\dd d/2)$ such that $i_{\star}+h_1, |s|+1+j_{\star}+h_2 \in D_S$, and $|h_1-h_2| \le \tau-1$. Without loss of generality, we assume $h_1\le h_2$, and the case of $h_1>h_2$ can be proved analogously by switching the roles of $s$ and $t$.  Now we consider two cases:
	\begin{itemize}
		\item \textbf{Case 1:} $\per(s[i_{\star}+h_1+\tau \dd i_{\star}+h_1+ 4\tau-2])>\tau/3$. 
		
		 Then, by the density condition of the $\tau$-synchronizing set $A$, we know that $A\cap [i_{\star}+h_1+\tau\dd i_{\star}+h_1+2\tau)$ is an non-empty set, and let $a$ be an arbitrary element of this set.
		Let $b = a-i_{\star}+j_{\star}$. Since $s[a\dd a+2\tau) = t[b\dd b+2\tau)$, or equivalently $S[a\dd a+2\tau)=S[|s|+1+b\dd |s|+1+b+2\tau)$, we have $|s|+1+b \in A$ by the consistency condition of $A$. 
		
		Note that we have 
		\begin{align*}
	b &=  j_{\star}+ h_2 +(a-i_{\star}-h_1)  -(h_2-h_1) \\ &\in [ j_{\star}+h_2+\tau - (h_2-h_1) \dd j_{\star}+h_2+ 2\tau -(h_2-h_1)) \\ & \subseteq [j_{\star}+h_2+1 \dd j_{\star}+h_2+2\tau),
		\end{align*}
		so $|s|+1+b\in A \cap [|s|+1+j_{\star}+h_2+1 \dd |s|+1+ j_{\star}+h_2+2\tau)$.
		
		 From the sparsity property of $A$ (\cref{lem:kk-property}), using Markov's inequality and a union bound, we can show that $|A \cap [i_{\star}+h_1\dd i_{\star}+h_1+2\tau)|\le 1000$ and $|A \cap [|s|+1+j_{\star}+h_2\dd |s|+1+ j_{\star}+h_2+2\tau)|\le 1000$ hold simultaneously with probability at least $1- 2\cdot 80/1000 >0.8$.
		 In this case, in Step 1 of \cref{defn:anchor} we must have $a\in L_{i_{\star}+h_1}$, and $|s|+1+b\in L_{|s|+1+j_{\star}+h_2}$. Then, setting $i=i_{\star},j=j_{\star},h = a-i_{\star}$ satisfies the requirement.
		\item \textbf{Case 2:} $p=\per(s[i_{\star}+h_1+\tau \dd i_{\star}+h_1+ 4\tau-2])\le \tau/3$.
		
		Then we have $ \per(s[i_{\star}+h_1+\tau \dd i_{\star}+h_1+3\tau-2])=p$. From $s[i_\star \dd i_\star+d)=t[j_\star\dd j_\star+d)$ and $0\le h_2-h_1\le \tau-1$, we also have $\per(t[j_{\star}+h_2+\tau \dd j_{\star}+h_2+3\tau-2])=p$. Hence, for both $L_{i_{\star}+h_1}$ and $L_{|s|+1+j_{\star}+h_2}$, we triggered Step 2 in \Cref{defn:anchor}. Then, we consider three subcases.
			\begin{itemize}
				\item \textbf{Case 2(i):} $\per(s[i_{\star} + h_1+\tau \dd i_{\star}+d))\neq p$. 
				
				In this case, the period $p$ of $s[i_{\star}+h_1+\tau \dd i_{\star}+h_1+3\tau-2]$ does not extend to its superstring $s[i_{\star} + h_1+\tau \dd i_{\star}+d)$, so the right boundary
				\[
	        r_s := \max\big \{r:  \per(s[i_\star+h_1+\tau\dd r])=p\big \}
	        \]
			must satisfy $i_\star+h_1+4\tau-2 \le r_s <i_\star+d-1$. Here we observe that $r_s$ is the same as the right boundary $r$ in Step 2 of \Cref{defn:anchor} for constructing $L_{i_\star+h_1}$.
			
			Let $r_t := r_s - i_\star + j_\star$. Then $j_\star+h_2+\tau < r_t$, and $t[j_\star+h_2+\tau \dd r_t+1] = s[i_\star+ h_2+\tau\dd r_s+1]$.
			Then from the definition of $r_s$, we can observe that 
	        \[r_t = \max\big \{r:  \per(t[ j_\star+h_2+\tau\dd r])=p\big \},\]
	        and $|s|+1+r_t$ must be the same as the right boundary $r$ in Step 2 of \Cref{defn:anchor} for constructing $L_{|s|+1+j_\star+h_2}$. 
	        
	        Let $P$ denote the Lyndon root of $s[i_\star+h_1+\tau\dd r_s]$, and let $P=s[i^{(e)}\dd i^{(e)}+p)=t[j^{(e)}\dd j^{(e)}+p)$ be the last occurrences of $P$ in $s[i_\star+h_1+\tau\dd r_s]$ and $t[j_\star+h_2+\tau\dd r_t]$. We must have $r_s-i^{(e)}=r_t-j^{(e)}$. Note that $i^{(e)}\in L_{i_\star+h_1}$ and $|s|+1+j^{(e)} \in L_{|s|+1+j_\star+h_2}$. So setting $i=i_\star,j=j_\star, h = i^{(e)}-i_\star$ satisfies the requirement.
	        
				\item \textbf{Case 2(ii):} $\per(s[i_{\star} + h_1+\tau \dd i_{\star}+d))= p$, but $\per(s[i_\star\dd i_\star+d))\neq p$. 
				
				In this case, the period $p$ fully extends to the right but not to the left. Using a similar argument as in Case 2(i), we can show that the left boundaries 
				\begin{align*}
				    \ell_s&:= \min \{\ell: \per(s[\ell\dd i_\star+h_1+3\tau-2])=p\},\\
				    \ell_t&:= \min \{\ell: \per(t[\ell\dd j_\star+h_2+3\tau-2])=p\}
				\end{align*}
				must satisfy  $\ell_s-i_\star = \ell_t-j_\star\ge 1$, and $\ell_s,|s|+1+\ell_t$ are the right boundaries in Step 2 of \Cref{defn:anchor} for constructing $L_{i_\star+h_1},L_{|s|+1+j_\star+h_2}$ respectively. Then, letting $s[i^{(b)}\dd i^{(b)}+p) = t[j^{(b)}\dd j^{(b)}+p)$ be the first occurrences of the Lyndon root in $s[\ell_s\dd i_\star+h_1+3\tau-2]$ and $t[\ell_t \dd j_\star+h_2+3\tau-2]$, we can similarly see that setting $i=i_\star,j=j_\star, h=i^{(b)}-i_\star$ satisfies the requirement.
				
				\item \textbf{Case 2(iii):}  $\per(s[i_\star\dd i_\star+d))= p$. 
				
				Let 
				\begin{align*}
				    \ell_s &:= \min\big \{\ell: \ell\ge \min \{1,i_\star+h_1-d\} \land \per(s[\ell\dd i_\star+h_1+3\tau-2])=p\big \},\\
				    \ell_t &:= \min\big \{\ell: \ell\ge \min \{1,j_\star+h_2-d\} \land \per(s[\ell\dd j_\star+h_2+3\tau-2])=p\big \}.
				\end{align*}
				Then $\ell_s,|s|+1+\ell_t$ are the right boundaries in Step 2 of \Cref{defn:anchor} for constructing the sets $L_{i_\star+h_1},L_{|s|+1+j_\star+h_2}$ respectively. We must have $\ell_s<i_\star$ and $\ell_t<j_\star$.
				
				Let $P$ be the Lyndon root of $s[i_\star\dd i_\star+d)$ and let $s[i^{(b)}\dd i^{(b)}+p)=t[j^{(b)}\dd j^{(b)}+p) = s[i'\dd i'+p)$ be the first occurrences of $P$ in $s[\ell_s\dd i_\star+d),t[\ell_t\dd j_\star+d)$, and $s[i_\star\dd i_\star+d)$ respectively. Let $i_{\star}' =i^{(b)}- (i'-i_\star),j_{\star}' =j^{(b)}- (j'-j_\star)$. Then, we have $s[i_\star\dd i_\star+d) = s[i_\star'\dd i_\star'+d)$ or $s[i_\star\dd i_\star+d) =s[i_\star'+p\dd i_\star'+p+d)$. Similarly we have $t[j_\star\dd j_\star+d) = t[j_\star'\dd j_\star'+d)$ or $t[j_\star\dd j_\star+d) =t[j_\star'+p\dd j_\star'+p+d)$. Note that $i^{(b)}, i^{(b)}+p \in L_{i_\star+h_1}$, and $|s|+1+j^{(b)}, |s|+1+j^{(b)}+p \in L_{|s|+1+j_\star+h_2}$. Hence, setting $i=i'_\star$ (or $i=i'_\star+p$), $j=j'_\star$ (or $j=j'_\star+p$), $h= i'-i_\star$ satisfies the requirement. 
			\end{itemize}
	\end{itemize}
	Hence, the desired $i,j$ and $h$ always exist.
\end{proof}

Finally, \cref{lem:good-anchor-} immediately follows from \cref{defn:anchor}, \cref{prop:local-access}, and \cref{lem:good-anchor}, by setting $\tau = \Theta(\sqrt{d})$.
\begin{comment}
Our quantum walk algorithm is almost the same as the warm-up algorithm (\Cref{sec:lcs-warm-up}), with the following two changes:
\begin{itemize}
    \item We use our newly defined good anchor set $L$ in place of the set $U$ used in the warm-up algorithm, so our Johnson graph has vertex set $\binom{L}{r}$.
    \item In each update step of the quantum walk, it takes $\tilde O(\tau+\sqrt{d})$ quantum queries to access an element in $L$, by \Cref{prop:local-access}. Whereas in the warm-up algorithm, it did not take any queries to  access the elements in $U$.
\end{itemize}
We set $\tau = \Theta(\sqrt{d})$, so that the update cost of the quantum walk is still $\tilde O(\sqrt{n})$.
By \Cref{prop:local-access} and \Cref{defn:dlcover}, $|L| \le 1003\cdot  |D_S| = O(n/\sqrt{d\tau}) =O(n/d^{3/4})$. Then, the query complexity (up to polylog factors) of the quantum walk is 
\begin{align}
&    \sfS + \sqrt{\frac{|L|^2}{r^2}}\cdot (\sfC + \sqrt{r} \cdot \sfU)\\
= \ & r\cdot \sqrt{d} + \frac{n/d^{3/4}}{r} \sqrt{r}\sqrt{d}\nonumber\\
 = \ & n^{2/3}.\nonumber
\end{align}
\end{comment}

\section{Minimal String Rotation}
\label{sec:lmsr}

\subsection{Minimal Length-\texorpdfstring{$\ell$}{l} Substrings}

Rather than work with the Minimal String Rotation problem directly, we present an algorithm for the following problem, which is more amenable to work with using our divide-and-conquer approach.

\defproblem{Minimal Length-$\ell$ Substrings}
{A string $s[1\dd n]$ and an integer $n/2\le \ell \le n$}
{Output all elements in $\arg\min_{1\le i\le n-\ell+1} s[i\dd i+l)$ represented as an arithmetic progression.}

The elements in the output are guaranteed to be an arithmetic progression thanks to \cref{lm:period-patterns}.

We will prove the following theorem.

\begin{thm}
\label{thm:minl}
Minimal Length-$\ell$ Substrings can be solved by a quantum algorithm with $n^{1/2+o(1)}$ query complexity and time complexity.
\end{thm}

For convenience, we also introduce the following problem.

\defproblem{Maximal String Rotation}
{A string $s$}
{Output a position $i\in [1\dd |s|]$ such that $s[j\dd |s|]s[1\dd j-1]\preceq s[i\dd |s|]s[1\dd i-1]$ holds for all $j\in [1\dd |s|]$. If there are multiple solutions, output the smallest such $i$. }

We now use a series of simple folklore reductions to show that the Minimal Length-$\ell$ Substrings problem generalizes the Minimal String Rotation problem.

\begin{prop}
\label[proposition]{prop:maxmin}
The Minimal String Rotation problem reduces to the Maximal String Rotation problem.
\end{prop}
\begin{proof}
Take an instance of the Minimal String Rotation problem, consisting of a string $s$ over an alphabet $\Sigma$, which recall we identify with the set $[1\dd |\Sigma|]$.
Consider the map $\varphi \colon \Sigma\to\Sigma$ defined by taking
  \[\varphi(c) = |\Sigma| - c + 1\]
for each character $c\in\Sigma$.
Let 
    \[t = \varphi(s[1])\cdots \varphi(s[n])\]
be the result of applying this map to each character of $s$.

By construction, for any $c,c'\in \Sigma$ we have $\varphi(c) \prec \varphi(c')$ if and only if $c' \prec c$.
Combining this observation together with the definition of lexicographic order, we deduce that for any indices $j,k\in [1\dd n]$ we have 
\[t[j\dd n]t[1\dd j-1]\preceq t[k\dd n]t[1\dd k-1]\]
if and only if
\[s[k\dd n]s[1\dd k-1]\preceq s[j\dd n]s[1\dd j-1].\]

Thus the solution to the Maximal String Rotation problem on $t$ recovers the solution to the Minimal String Rotation problem on $s$, which proves the desired result.
\end{proof}

\begin{prop}
\label[proposition]{rot-suff}
The Maximal String Rotation problem reduces to the Maximal Suffix problem.
\end{prop}
\begin{proof}
Take an instance of the Maximal String Rotation problem, consisting of a string $s$ of length $n$.

Let $t = ss$
be the string of length $2n$ formed by concatenating $s$ with itself.
Suppose $i$ is the starting index of the maximal rotation of $s$.
Then we claim that $i$ is the starting index of the maximal suffix of $t$ as well.

Indeed, take any position $j\in [1\dd 2n]$ in string $t$ with $j\neq i$.

If $j > n$, then we can write $j = n + \Delta$ for some positive integer $\Delta \le n$.
In this case we have 
    \[t[j\dd 2n] \prec t[\Delta\dd 2n] \]
because the string on the left hand side is a proper prefix of the string on the right hand side.
Thus $j$ cannot be the starting position of a maximal suffix for $t$.

Otherwise, $j\le n$.
Note that we can write
    \begin{equation}
    \label{t1}
    t[i\dd 2n] = s[i\dd n]s[1\dd i-1] s[i\dd n] \quad \text{and} \quad 
    t[j\dd 2n] = s[j\dd n]s[1\dd j-1] s[j\dd n].
    \end{equation}
    
Since $i$ is a solution to the Maximal String Rotation problem, we know that either
    \[ s[j\dd n]s[1\dd j-1]\prec s[i\dd n]s[1\dd i-1] \]
or $ s[j\dd n]s[1\dd j-1] = s[i\dd n]s[1\dd i-1]$ and $i<j$.

In the first case, the decompositions from \cref{t1} immediately imply that 
    \[t[j\dd 2n] \prec t[i\dd 2n]\]
by considering the length $n$ prefixes of the two strings.
In the second case, since $ s[j\dd n]s[1\dd j-1] = s[i\dd n]s[1\dd i-1]$ and $i<j$ the decompositions from \cref{t1} imply that
    \[t[j\dd 2n] \prec t[i\dd 2n]\]
because the string on the left hand side is a proper prefix of the string on the right hand side.
Combining these results, we see that the solution to the Maximal Suffix problem on $t$ is the index $i$ which solves the Maximal String Rotation problem on $s$.
\end{proof}

\begin{prop}
\label[proposition]{suff:maxmin}
The Maximal Suffix problem reduces to the Minimal Suffix problem.
\end{prop}
\begin{proof}
Take an instance of the Maximal Suffix problem, consisting of a string $s$ of length $n$ over an alphabet $\Sigma = [1\dd |\Sigma|]$.
Let $\sigma = |\Sigma|+1$ denote a character lexicographically after all the characters in $\Sigma$.
As in the proof of \cref{prop:maxmin}, consider the map $\varphi \colon \Sigma\to\Sigma$ defined by taking
  \[\varphi(c) = |\Sigma| - c + 1\]
for each character $c\in\Sigma$.
Now, build the string
    \[t = \varphi(s[1])\varphi(s[2])\cdots \varphi(s[n])\sigma\]
formed by applying $\varphi$ to each character of $s$ and then appending $\sigma$ to the end.

Suppose that $s[i\dd n]$ is the maximal suffix of $s$.
We claim that $t[i\dd n+1]$ is the minimal suffix of $t$.
Thus solving the Minimal Suffix problem on $t$ recovers a solution to the Maximal Suffix problem on $s$.

To see this, note that take any index $1\le j\le n$ with $j\neq i$.
By assumption
    \begin{equation}
    \label{eq:prec}
    s[j\dd n] \prec s[i\dd n].
    \end{equation}
This can happen one of two ways.

First, it could be that $j > i$ and the string on the left hand side above is a proper prefix of the string on the right hand side.
In this case we must have 
    \[t[i\dd n+1] = \varphi(s[i])\cdots \varphi(s[n])\sigma \prec \varphi(s[j])\cdots \varphi(s[n])\sigma = t[j\dd n+1]\]
because the string on the left hand side agrees with the string on the right hand side for the first $j$ positions, but then at the $(n-j+2)^{\text{th}}$ position, the string on the right hand side has the character $\sigma$, which is larger than the corresponding character $\varphi(s[n - (j-i)])$ from the string on the left hand side.

Otherwise, \cref{eq:prec} holds because there exists some nonnegative integer $\Delta$ such that $s[j+\Delta] < s[i+\Delta]$ and $s[j+d] = s[i+d]$ for all nonnegative $d<\Delta$.
By definition, $\varphi(c)\prec \varphi(c')$ if and only if characters $c' \prec c$ for all $c,c'\in \Sigma$.
Thus in this case too we have 
    \[t[i\dd n+1] = \varphi(s[i])\cdots \varphi(s[n])\sigma \prec \varphi(s[j])\cdots \varphi(s[n])\sigma = t[j\dd n+1]\]
because the strings agree for the first $\Delta$ characters, but then at the $(\Delta+1)^{\text{st}}$ position, the string on the right hand side has the character $\varphi(s[j+\Delta])$, which is larger than the corresponding character $\varphi(s[i+\Delta])$ from the string on the left hand side by our observation on $\varphi$.
Finally, note that the suffix $t[n+1] = \sigma$ is larger than every other suffix of $t$ by construction, and is thus not a minimal suffix of $t$.
Thus the minimal suffix of $t$ corresponds to the maximal suffix of $s$, and the  reduction is correct.
\end{proof}

\begin{prop}
\label[proposition]{fin:red}
The Minimal Suffix problem reduces to the Minimal Length-$\ell$ Substrings problem.
\end{prop}
\begin{proof}
Take an instance of the Minimal Suffix problem, consisting of a string $s$ of length $n$.
Consider the string of length $2n-1$ of the form
    \[t = s\underbrace{00\cdots 0}_{n-1\text{ times}}\]
formed by appending $n-1$ copies of a character $0$, smaller than every character from the alphabet $\Sigma$ of $s$, to the end of $s$.

Let $i$ be the starting index of the minimal suffix of $s$.
We claim that $i$ is also the unique index returned by solving the Minimal Length-$n$ Substrings problem on $t$ (note that $n$ is at least half the length of $t$).

Indeed, take any index $j\in [1\dd n]$ with $j\neq i$.
By assumption we have 
    \[s[i\dd n] \prec s[j\dd n].\]
Because the string on the left hand side occurs strictly before the string on the right hand side in lexicographic order, appending any number $0$s to the ends of the strings above cannot change their relative order.
Thus
    \[t[i\dd i+n) = s[i\dd n] \underbrace{00\cdots 0}_{i-1\text{ times}}
     \prec s[j\dd n] \underbrace{00\cdots 0}_{j-1\text{ times}} = t[j\dd j+n)\]
as well.
Because this holds for all $j\neq i$ we get that $i$ is the unique position output by solving the Minimal Length-$n$ Substrings problem on $t$.
This proves the reduction is correct.
\end{proof}

By chaining the above reductions together, we obtain the following corollary of \cref{thm:minl}.
\begin{thm}
\label{thm:lexico-main}
Minimal String Rotation, Maximal Suffix, and Minimal Suffix can be solved by a quantum algorithm with $n^{1/2+o(1)}$ query complexity and time complexity. 
\end{thm}
\begin{proof}
    By combining the results of \Cref{prop:maxmin}, \Cref{rot-suff}, \Cref{suff:maxmin}, and \Cref{fin:red}, we see that all the problems mentioned in the theorem statement reduce to the Minimal Length-$\ell$ Substrings problem. 
    Each of the reductions only involves simple substitutions and insertions to the input strings.
    
    In particular, by inspecting the proofs of the propositions, we can verify that for an input string $s$ and its image $t$ under any of these reductions, any query to a character of $t$ can be simulated with $O(1)$ queries to the characters of $s$.
    Thus, we can get a $n^{1/2 + o(1)}$ query and time quantum algorithm for each of the listed problems by using the algorithm of \Cref{thm:minl} and simulating the aforementioned reductions appropriately in the query model.
\end{proof}

\begin{remark}
We remark that, from the $\Omega(\sqrt{n})$ quantum query lower bound for Minimal String Rotation \cite{ying}, this chain of reductions also implies that Maximal Suffix and Minimal Suffix require $ \Omega(\sqrt{n})$ quantum query complexity.
\end{remark}
It remains to prove \cref{thm:minl}.  
To solve the Minimal Length-$\ell$ Substrings problem, it suffices to find any individual solution
\[i \in \displaystyle \argmin_{1\le i\le n-\ell +1} s[i\dd i+\ell ),\] and then use the quantum Exact String Matching algorithm to find all the elements (represented as an arithmetic progression) in $\tilde O(\sqrt{n})$ time.
Our approach will invoke the following ``exclusion rule,'' which simplifies the previous approach used in \cite{ying}.
We remark that similar kinds of exclusion rules have been applied previously in parallel algorithms for Exact String Matching \cite{DBLP:journals/siamcomp/Vishkin91} and Minimal String Rotation \cite{DBLP:journals/tcs/IliopoulosS92} (under the name of ``Ricochet Property'' or ``duel''), as well as the quantum algorithm by Wang and Ying \cite[Lemma 5.1]{ying}. 
The advantage of our exclusion rule is that it naturally yields a \emph{recursive} approach for solving the Minimal Length-$\ell$ Substrings problem.

\begin{lem}[Exclusion Rule]
\label[lemma]{exclusion}
In the Minimal Length-$\ell$ Substrings problem with input $s[1\dd n]$ with $n/2\le \ell \le  n$, let 
\[I:= \argmin_{1\le i\le n-\ell+1} s[i\dd i+\ell)\]
denote the set of answers forming an arithmetic progression.
For integers $a\ge 1,k\ge 1$ such that $a+k\le n-\ell+1$, let $J$ denote the set of answers in  the Minimal Length-$k$ Substrings problem on the input string $s[a\dd a+2k)$. 
Then if $\{\min J,\max J\} \cap  I = \emptyset$, we must have $J\cap I = \emptyset$.
\end{lem}

\begin{proof}
First observe that 
\[a+2k-1 \le  n-\ell+k \le 2(n-\ell) \le  n,\]
so $s[a\dd a+2k)$ is a length-$2k$ substring of $s$.
Since the statement is trivial for $|J|\le 2$, we assume $J$ consists of $j_1<j_2<\dots <j_m$ where $m\ge 3$.  Let $p = j_2-j_1$. Then $p = (j_m-j_1)/(m-1) \le k/2$. 
Then from 
\[s[j_1\dd j_1+k)= s[j_2\dd j_2+k)=\cdots = s[j_m\dd j_m+k)\]
we know that $p$ must be a period of $s[j_1\dd j_m+k)$.\footnote{In fact, $p$ is the minimum period of this substring.}
We consider the first position $r$ where this period $p$ stops, that is, $r:= \min \{j_m+k \le r \le n : s[r]\neq s[r-p]\}$. If such $r$ does not exist, let $r=n+1$. 
With this setup, we now proceed to prove the contrapositive of the original claim.  

Suppose $j_q \in I$ for some $1\le q\le m$. We consider three cases.
\begin{itemize}
    \item \textbf{Case 1:} $r\ge j_m+\ell$. 
    
    In this case, we must have $s[j_1\dd j_1+\ell) = s[j_2\dd j_2+\ell) = \cdots = s[j_m\dd j_m+\ell)$. Then, $j_q\in I$ implies $j_1\in I$ and $j_2\in I$.
    \item \textbf{Case 2:} $r< j_m+\ell$, and $s[r]<s[r-p]$.
    
    For every $1\le t\le m-1$, by the definition of $r$, we must have $s[j_{t+1}\dd r) = s[j_t\dd r-p)$. Then from $s[r]<s[r-p]$ we have $s[j_t\dd j_t+\ell)\succeq s[j_{t+1}\dd j_{t+1}+\ell)$. Hence, $j_q \in I$ implies $j_{q+1},j_{q+1},\dots,j_m \in I$.
    
    \item \textbf{Case 3:} $r<j_m+\ell$, and $s[r]>s[r-p]$.
    
    By an argument similar to Case 2, we can show $s[j_t\dd j_t+\ell)\preceq s[j_{t+1}\dd j_{t+1}+\ell)$. Then, $j_q \in I$ implies $j_{q-1},j_{q-2},\dots,j_1 \in I$.
\end{itemize}
Thus $\{j_1,j_m\} \cap I \neq \emptyset$ in all of the cases, which proves the desired result.
\end{proof}

\subsection{Divide and Conquer Algorithm}

To motivate our quantum algorithm, we first describe a classical algorithm for the Minimal $n/2$-length Substring problem which runs in $O(n\log n)$ time (note that other classical algorithms can solve the problem faster in $O(n)$ time).
Our quantum algorithm will use the same setup, but obtain a speed-up via Grover search.
For the purpose of this overview, we assume $n$ is a power of $2$.
The classical algorithm works as follows:

Suppose we are given an input string $s$ of length $n$ and target substring size $\ell = n/2$.
Set $m = \ell/2 = n/4$.
Then the half of the solution (i.e. the first $m$ characters of a minimum length $\ell$-substring) are contained entirely in either the block $s_1 = s[1\dd n/2]$ or the block $s_2 = s[n/4\dd 3n/4)$.

With that in mind, we recursively solve the problem on the strings $s_1$ and $s_2$ with target size $m$ in both cases. 
Let $u_1$ and $v_1$ be the smallest and largest starting positions returned by the recursive call to $s_1$ respectively.
Define $u_2$ and $v_2$ as the analogous positions returned by the recursive call to $s_2$.
Then by \Cref{exclusion}, the true starting position of the minimal $\ell$-length substring of $s$ is in $\set{u_1, u_2, v_1, v_2}$.

We identify the $\ell$-length substrings starting at each of these positions, and find their lexicographic minimum in $O(n)$ time via linear-time string comparison.
This lets us find at least one occurrence of the minimum substring of length $\ell$.
Then, to find all occurrences of this minimum substring, we use a linear time string matching algorithm (such as the classic Knuth-Morris-Pratt algorithm \cite{DBLP:journals/siamcomp/KnuthMP77}) to find the first two occurrences of the minimum length $\ell$ substring in $s$.
The difference between the starting positions  then lets us determine the common difference of the arithmetic sequence of positions encoding all starting positions of the minimum substring.

If we let $T(n)$ denote the runtime of this algorithm, the recursion above yields a recurrence
\[T(n) = 2T(n/2) + O(n)\]
which solves to  $T(n) = O(n\log n)$.

\subsection{Quantum Speedup}

Next, we show how to improve the runtime of this divide-and-conquer approach in the quantum setting.
The key change is to break the string into $b$ blocks, and apply quantum minimum finding over these blocks which only takes $\tilde O(\sqrt{b})$ recursive calls, instead of $b$ recursive calls needed by the classical algorithm.  We will set $b$ large enough to get a quantum speedup.

\begin{proof} (of \Cref{thm:minl})
Let $b$ be some parameter to be set later.
For convenience assume that $b$ divides both $\ell$ and $n$ (this assumption does not affect the validity of our arguments, and is only used to let us avoid working with floor and ceiling functions).
Set $m = \ell/b$.

For each nonnegative integer $k\le \lfloor n/m\rfloor -2$ we define the substring
    \[s_k = s(km\dd (k+2) m].\]
Also set $s_{\lfloor n/m\rfloor -1} = s(n-2m\dd n]$.

These $s_k$ blocks  each have length $2m$, and together cover every substring of length $m$ in $s$.
Let $P$ be the minimum length-$\ell$ substring in $s$.
By construction, the first $m = \ell/b$ characters of $P$ is contained entirely in one of the $s_k$ blocks. 

For each block $s_k$, let $P_k$ denote its minimum length-$m$ substring and let $u_k$ and $v_k$ be the smallest and largest starting positions respectively of an occurrence of $P_k$ in $s_k$.
The lexicographically smallest prefix $P_k$ will make up the first $m$ characters of the minimum length-$\ell$ substring.
Thus by \Cref{exclusion}, we know the minimum length-$\ell$ substring of $s$ must start at position $u_k$ or $v_k$ for some index $k$.

We now use quantum minimum finding to find $P$. 
We search over the $\Theta(n/m) = \Theta(b)$ blocks above.
To compare blocks $s_i$ and $s_j$, we recursively solve the Minimal Length-$m$ Substrings problem on $s_i$ and $s_j$ to find positions $u_i, v_i$ and $u_j, v_j$.
Then we look at the substrings of length $\ell$ starting at these four positions. 
By binary search and Grover search (\cref{lem:lcp-grover}), in $\tilde{O}(\sqrt n)$ time we can determine which of these four substrings is lexicographically the smallest.
If the smallest of these substrings came from $s_i$ we say block $s_i$ is smaller than block $s_j$, and vice versa.

After running the minimum finding algorithm, we will have found $P$.
To return all occurrences of $P$, we can then use the quantum algorithm for Exact String Matching to find the two leftmost occurrences and the rightmost occurrence of $P$ in $s$ in $\tilde{O}(\sqrt n)$ time. 
Together they determine the positions of all copies of $P$ in $s$ as an arithmetic sequence, which we can return to solve the original problem.
%By computing the distance between the starting positions of these two copies of $P$, we will get the value of the the common difference for the arithmetic sequence of all starting positions of copies of $P$ in $s$.
%We can also find the rightmost occurrence of $P$ in $s$ by solving Exact String Matching again in the same amount of time to find the final term of this arithmetic sequence.

It remains to check the runtime of the algorithm.
Let $T(n)$ denote the runtime of the algorithm with error probability at most $1/n$.
Recall that our algorithm solves Minimum Finding over $\Theta(b)$ blocks, where each comparison involves a recursive call on strings of size $2m = \Theta(n/b)$ and a constant number of string comparisons  of length $n$ (via \cref{lem:lcp-grover}), and finally solves  Exact String Matching for strings of size $\Theta(n)$.
Hence we have the recurrence (assuming all logarithms are base 2)
    \[T(n) \le \tilde O(\sqrt{b})\cdot \left ( T(n/b) + \tilde O(\sqrt{n})\right ) + \tilde O(\sqrt{n}) = c(\log b)^{e}\sqrt{b}\left(T(n/b) + \sqrt{n}\right)\]
for some constants $c, e  > 0$, where the polylogarithmic factors are inherited from the subroutines we use and the possibility of repeating our steps  $O(\log n)$ times to drive down the error probability.
Now set 
\[b= 2^{d(\log n)^{2/3}}\] 
for some constant  $d$.
We claim that for sufficiently large $d$, we recover a runtime of $T(n) = n^{1/2} \cdot 2^{d(\log n)^{2/3}}$.

We prove this by induction.
The result holds when $n$ is a small constant by taking $d$ large enough.
Now, suppose we want to prove the result for some arbitrary $n$, and that the claimed runtime bound holds on inputs of size less than $n$.
Then using the recurrence above and the inductive hypothesis we have
    \begin{align*}
        T(n)&\le  c(\log b)^{e}\sqrt{b}\left(T(n/b) + \sqrt{n}\right) \\ &\le  c(\log b)^{e}\sqrt{n} \left(2^{d(\log(n/b))^{2/3}} + \sqrt b\right)\\&\le  2c(\log b)^{e}\sqrt{n} \cdot 2^{d(\log(n/b))^{2/3}}, 
    \end{align*}
 where the last inequality follows from $d(\log (n/b))^{2/3} \ge d(\log (\sqrt{n}))^{2/3} >  \frac{1}{2} d(\log n)^{2/3} = \log (\sqrt{b})$ for large enough $n$.
Equivalently, this means that 
    \begin{equation}
    \label{eq:frac-bound}
    \frac{T(n)}{n^{1/2}2^{d(\log n)^{2/3}}}
    \le 
    2c \cdot 2^{e(\log\log b) -d\left((\log n)^{2/3} - 
    (\log n - \log b))^{2/3}\right)}.
    \end{equation}

Using the mean value theorem, we can bound
\begin{align*}
 (\log n)^{2/3} - (\log n - \log b)^{2/3}  & \ge (2/3) (\log b)(\log n)^{-1/3}\\
    & = (2/3)d(\log n)^{1/3} \\
    &\ge  \omega(\log \log b),
\end{align*}
where the last inequality follows from $\log \log b = \log d + (2/3)\log \log n$.
Thus, by taking $d$ to be a large enough constant in terms of $c$ and $e$, we can force the right hand side of \Cref{eq:frac-bound} to be less than $1$, which proves that 
    \[T(n) \le n^{1/2}2^{d(\log n)^{2/3}}.\]
This completes the induction, and proves that we can solve the Minimum Length-$\ell$ Substrings problem in the desired runtime as claimed.
\end{proof}

\subsection{Longest Lyndon Substring}
The technique we use to solve the Minimal String Rotation problem can also be adapted to get a quantum speed-up for solving the Longest Lyndon Substring problem.

\begin{thm}
\label{thm:long-lyndon}
The Longest Lyndon Substring problem can be solved by a quantum algorithm with $n^{1/2+o(1)}$ query complexity and time complexity. 
\end{thm}

A difficulty in solving Longest Lyndon Substring compared to other string problems such as LCS and Longest Palindromic Substring is that the lengths of Lyndon Substrings do not have the monotone property, and hence we cannot use binary search (the Longest Square Substring problem in \cref{sec:squared} also has the same issue). To overcome this issue, we first present a simple reduction. 

\begin{thm}
\label{thm:search}
For any constant $0<\eps <1$, suppose there is a $T(d)$-time quantum algorithm (where $T(d) \ge \Omega(\sqrt{d})$) for solving the Longest Lyndon Substring problem on string $s$ of length $|s|= (1+2\eps)d$ with the promise that the longest Lyndon substring of $s$ has length in the interval $[d, (1+\eps)\cdot d)$. And suppose there is an $T(d)$-time quantum algorithm for checking whether an $O(d)$-length string is a Lyndon word.

Then, there is an algorithm in time $\tilde O(T(n))$ for solving the Longest Lyndon Substring problem on length-$n$ strings in general case.
\end{thm}
\begin{proof}
Let $s$ be the input string of length $n$. For each nonnegative integer $i \le \lceil (\log n)/(\log (1+\eps))\rceil - 1$, we look for a longest Lyndon substring of $s$ whose length is in the interval $\left[(1+\eps)^i, (1+\eps)^{i+1}\right)$, and return the largest length (after certifying that it is indeed a Lyndon substring) found. This only blows up the total time complexity by an $O(\log n)$ factor.

For each $i$, we define the positions $j_k:= 1+ k\cdot \eps d/2$ for all $0\le k< 2n/(\eps d)$, 
and consider the substrings 
\[s[j_0\dd j_0+(1+2\eps)d), s[j_1\dd j_1+(1+2\eps)d),\dots\] 
Note that, if the longest Lyndon substring of $s$ has length in the interval $[d,(1+\eps)d)$, then it must be entirely covered by some of these substrings.  For each of these substrings, its longest Lyndon substring can be computed in $T(d)$-time by the assumption. Then, we use the quantum maximum finding algorithm  (see \cref{sec:primitive}) to find the longest among these $2n/(\eps d)$ answers , in $\tilde O(\sqrt{2n/(\eps d)} \cdot T(d)) = \tilde O(\sqrt{n}\cdot T(d)/\sqrt{d}) \le \tilde O(T(n))$ overall time, where we used the assumption of $T(d) \ge \Omega(\sqrt{d})$.
\end{proof}

Now, we are going to describe an $d^{1/2+o(1)}$-time quantum algorithm  for solving the Longest Lyndon Substring problem on string $s$ of length $|s|= (1+2\eps)d$, with the promise that the longest Lyndon substring of $s$ has length in the interval $[d, (1+\eps)\cdot d)$.  Combined with the reduction above, this proves \cref{thm:long-lyndon},  since a string is Lyndon if and only if its minimal suffix is itself (see \cref{sec:prelim:notation}) and can be checked by our Minimal Suffix algorithm. We will set $\eps = 0.1$.

We will make use of the following celebrated fact related to Lyndon substrings. 
\begin{definition}[Lyndon Factorization \cite{chen1958free,DBLP:journals/jal/Duval83}]
    Any string $s$ can be written as a concatenation
        \[s = s_1s_2\cdots s_k\]
    where each string $s_i$ is a Lyndon word, and $s_1  \succeq s_2  \succeq \dots  \succeq s_k$.
    This decomposition is unique, and called the \emph{Lyndon factorization} of $s$.
    The $s_i$ are called \emph{Lyndon factors} of $s$
\end{definition}
    
The following fact characterizes the longest Lyndon substring in a given string.
\begin{prop}[e.g., {\cite[Lemma 3]{DBLP:conf/cpm/UrabeNIBT18}}]
    The longest Lyndon substring of a string $s$ is necessarily a longest Lyndon factor of $s$.
\end{prop}

Then, given the promise about the input string $s$ of length $(1+2\eps)d$, we know $s$ has Lyndon factorization $s_1\cdots  s^* \cdots s_k$, where $|s^*| \in [d, (1+\eps)d)$. The remaining task is to identify the position and length of the Lyndon factor $s^*$.

\begin{lem}[The position of $s^*$]
Suppose $s[i\dd i+|s^*|) = s^*$. Then,
$s[i \dd |s|]$ must be the minimal suffix among all $i\in [1\dd \eps d + 1]$. 
\end{lem}
\begin{proof}
Note that $i\in [1\dd \eps d+1]$ due to $|s| = (1+2\eps) d $ and $|s^*| \in [d,(1+\eps)d)$.
For any other starting position $j \in [1\dd \eps d+1]$, we will prove that $s[j\dd |s|] \succ s[i\dd |s|]$. We consider two cases.

\paragraph*{Case 1:} $j>i$.
    In this case we must have $j \in (i\dd i+|s^*|)$ due to the length constraints. Then, we have $s[j\dd i+|s^*|) \succ s[i\dd i+|s^*|)$ due to the fact that $s^*$ is a Lyndon word, which immediately implies $s[j\dd |s|] \succ s[i\dd |s|]$, since $\left \lvert s[i\dd i+|s^*|)\right \rvert > \left \lvert s[j\dd i+|s^*|)\right \rvert$.
    
   \paragraph*{Case 2:} $j<i$. 
    Then, suppose a Lyndon factor $s_t$ prior to $s^*$ occurs at $s_t = s[j'\dd j'']$ with $j'\le j\le j''$. Then, we have $s[j\dd j''] \succeq s[j'\dd j''] = s_t \succeq s^*$ by the property of Lyndon factorization. Then, from the length constraint $|s[j\dd j'']|< |s^*|$, we necessarily have  $s[j\dd |s|]\succ s[i\dd |s|] $.  
\end{proof}
Then we can find the starting position of $s^*$, by looking for the minimal suffix of $s$ whose starting position is in $[1\dd \eps d+1 ]$. We observe that, this task can be reduced to the the Minimum Length-$\ell$ Substrings problem using the same reduction as in \cref{fin:red} by appropriately adjusting the lengths, and we omit the proof here.  Hence, we can find the starting position of $s^*$ in $d^{1/2+o(1)}$ time.

We can now without loss of generality assume that $s^*$ appears as the first Lyndon factor of the input string $s= s^*s_2s_3\cdots s_m$ of length $|s|\le (1+2\eps)d$. It remains to find the ending position of $s^*$. We need the following definition.
\begin{definition}
We say a string $s$ is \emph{pre-Lyndon}, if there is a Lyndon word $t$ such that $s$ is a prefix of $t$.
\end{definition}
We have the following characterization of pre-Lyndon strings.
\begin{prop}[e.g., {\cite[Lemma 10]{DBLP:conf/cpm/UrabeNIBT18}}]
For any pre-Lyndon string $w$, there exists a unique Lyndon word $x$  such that $w=x^kx'$ where $k\ge 1$, and $x' = x[1\dd i]$ for some $i \in [0\dd |x|-1]$. Here $x^k$ denotes concatenating $x$ for $k$ times.
\label[proposition]{prop:prelyn}
\end{prop}

Note that we can check whether a string $w$ is pre-Lyndon, in $|w|^{1/2+o(1)}$ time.
\begin{lem}
Given any string $w$ of length $d$, we can check whether it is pre-Lyndon in $d^{1/2+o(1)}$ quantum time. Moreover, if $w$ is pre-Lyndon, we can find its decomposition described in \cref{prop:prelyn} also in $d^{1/2+o(1)}$ quantum time.
\label[lemma]{lem:computeprelyndon}
\end{lem}

\begin{proof}
 We assume $w$ is indeed a pre-Lyndon string, and has decomposition $w=x^kx'$ as described in \cref{prop:prelyn}. 

We first observe that, the minimal rotation of $w=x^kx'$ must equal $x'x^k$. This observation can be easily proved from the fact that the Lyndon word $x$ is strictly smaller than all other rotations of $x$. In the case of $|x'|\ge 1$, we can compute the shift of the minimal rotation of $w$ (which must be unique), and obtain the length $|x'|$. We can also detct the case of $|x'|=0$, by finding that $w$ itself equals the minimal rotation of $w$.

After finding $|x'|$, we are left with the part $w'=x^k$, and we know that $|x|=\per(|w'|)$. We can then compute $\per(|w'|)$ by finding the second earliest occurrence of $w'$ in the string $w'w'$, using the quantum Exact String Matching algorithm \cite{matching}. (Alternatively, we can use the periodicity algorithm of Wang and Ying \cite{ying}) 

Finally, after obtaining $x$ and $x'$, we certify that $w$ is indeed a pre-Lyndon string, by checking that $x$ is a Lyndon word, $x'$ is a prefix of $x$, and $w=x^k x'$, in $\tilde O(\sqrt{|w|})$ time by Grover search.
\end{proof}

Then, on the input string $s$ with Lyndon factorization $s= s^*s_2s_3\cdots s_m$, we apply \cref{lem:computeprelyndon} with binary search to find the maximum position $i \in [|s|]$ such that $s[1\dd i]$ is pre-Lyndon. We must have $i\ge |s^*|$, by the definition of pre-Lyndon string and the fact that $s^*$ is Lyndon. We also obtain the decomposition of $s[1\dd i] = x^kx'$ described in \cref{prop:prelyn}, where $x$ is a Lyndon word with proper prefix $x'$. Note that the longest Lyndon prefix of $x^k x'$ must be $x$, since any longer prefix of $x^kx'$ can be written as $x^{j}x''$ and obviously has a smaller suffix than itself. Then, from the fact that $s^*$ is the longest Lyndon prefix of $s[1\dd i]$, we know $x=s^*$. Hence, we have completely determined $s^*$.

\begin{remark}
We can show that the Longest Lyndon Substring problem requires $\Omega(\sqrt{n})$ quantum queries, by a simple reduction from the unstructured search problem. Suppose we are given a string $s \in \{0,1\}^n$ and want to decide whether there exists $i\in [n]$ such that $s[i] = 1$. We create another string $s':=s0^n1$ by appending $n$ zeros and a one after $s$. Then, if $s=0^n$, then the longest Lyndon substring of $s'$ will be $s'$ itself. If there is at least a one in $s$, then $s'$ cannot be a Lyndon word, since its suffix $0^n1$ must be smaller than $s'$. Hence, by \cite{DBLP:journals/siamcomp/BennettBBV97}, this implies that Longest Lyndon Substring problem requires query complexity $\Omega(\sqrt{n})$.
\end{remark}

\section{Longest Square Substring}
\label{sec:squared}

Recall that in the Longest Square Substring problem, we are given a string $s$ of length $n$ and tasked with finding the largest positive integer $\Delta$ such that there exists some index $1\le i \le n-2\Delta + 1$ with $s[i\dd i+\Delta) = s[i+\Delta\dd i+2\Delta)$.
In other words, we want to find the maximum size $\ell = 2\Delta$ such that $s$ contains a $\Delta$-periodic substring of length $\ell$.
We call $\Delta$ the \emph{shift} and $\ell$ the \emph{length} of the longest square substring.
We refer to the substrings $s[i\dd i+\Delta)$ and $s[i+\Delta\dd i+2\Delta)$ as the \emph{first half} and \emph{second half} of the solution respectively.

In this section, we present a quantum algorithm which solves this problem on strings of length $n$ in $\tilde{O}(\sqrt n)$ time.
We follow this up with a brief argument indicating why this algorithm is optimal up to polylogarithmic factors in the query complexity.

\begin{thm}
\label{thm:lss-main}
The Longest Square Substring problem can be solved by a quantum algorithm with $\tilde O(\sqrt n)$ query complexity and time complexity.
\end{thm}

\begin{proof}
Let $s$ be the input string of length $n$.
    
Set $\eps = 1/10$.
For each nonnegative integer $i \le \lceil (\log n)/(\log (1+\eps))\rceil - 1$,
we look for a longest square substring of $s$ whose length is in the interval $\left[(1+\eps)^i, (1+\eps)^{i+1}\right)$.
We begin with $i$ equal to its upper bound, and then keep decrementing $i$ until we find a square substring in the relevant interval.
The first time we find such a string we return it and halt.
If we never find such a string we report that $s$ has no square substring.
We try out $O(\log n)$ values of $i$, so it suffices to solve each of these subproblems in $\tilde{O}(\sqrt n)$ time
(this is very similar to the argument used in \Cref{thm:search}).

If $s$ has no square substring, our algorithm will never find a solution a will correctly detect that there is none.
Hence in the remainder of this proof, suppose that $s$ contains a square substring, and let $\ell = 2\Delta$ be the length of the longest square substring in $s$.
Let $i$ be the unique positive integer with $\ell \in \left[(1+\eps)^i, (1+\eps)^{i+1}\right)$.
We will eventually reach this value of $i$ since we cannot have found any square substrings of larger size.
Write $d = (1+\eps)^i$ and, for convenience, assume that $d$ is an integer multiple of $10$ (this will not affect the correctness of the arguments below).

We start by sampling a uniform random position $g$ in the string.
We say $g$ is \emph{good} if there exists a square substring $A$ of size $\ell$ in $s$ with the property that $g$ is among the first $d/10$ positions in $A$.
Note that $g$ is good with probability at least $\Omega(d/n)$.

Suppose $g$ is a good position.
Now, consider the substring 
\[P = s(g\dd g+2d/5]\]
of length $2d/5$ starting immediately after $g$.
Since $g$ is good, the end of $P$ is at most the \[d/10 + 2d/5\le d/2 \le \ell/2\]
position character in $A$.
Hence $P$ is contained completely in the first half of $A$.

Now define $S = s(g+2d/5 \dd g+(1+\eps)d]$ to be an $O(d)$ length substring of $s$ starting immediately after $P$.
Since $g$ is good and $A$ has length at most $(1+\eps)d$, we know that the second half of $A$ is contained completely in $S$.
In particular, $S$ must contain at least one copy of $P$.

By solving the Exact String Matching problem with $S$ and $P$, we can find the leftmost and rightmost occurrences of $P$ in $S$ in $\tilde{O}(\sqrt d)$ time.
If these copies occur at the same position, $S$ actually contains only a single copy of $P$, and otherwise $S$ contains multiple copies of $P$.
We consider these cases separately. 

\paragraph*{Case 1: Unique Copy.}
Suppose $S$ has only one copy of $P$.
Let this copy begin after position $h$, so that $s(h \dd h+2d/5]=P.$
Thus we get that $\Delta=h-g$, since the shift of the longest square substring equals the distance between copies of $P$ in the first and second half of $A$.
Now, find the largest nonnegative integer  $j\le 2d/5$ such that $s(g-j\dd  g] = s(h-j\dd  h]$.
All we are doing in this step is extending the copies of $P$ backwards while keeping them identical (pictured in the second image of \cref{fig:lss-case1} as light blue rectangles). 
This takes $\tilde{O}(\sqrt d)$ time via Grover search. 

%% Figure LSS Case 1
\begin{figure}[t]
\centering
\def\svgwidth{\linewidth}
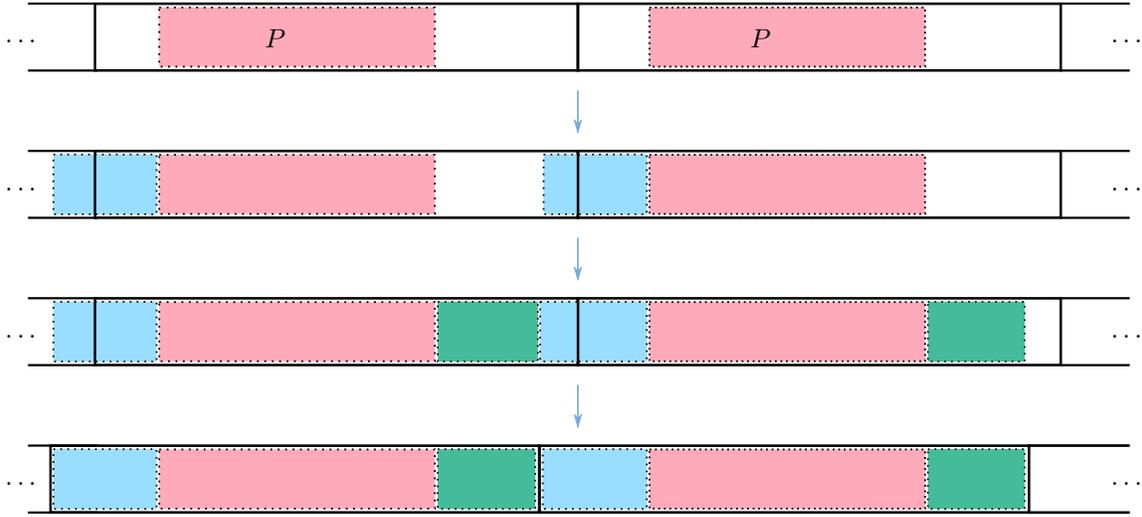
\caption{An example of how we extract a square substring when there is only one copy of $P$. 
The black vertical line segments in the top three images bound the first half and second half of the true optimal square substring $A$. 
When we extend the patterns backwards we cover the entire prefix, we are guaranteed to reach the boundaries of $A$, but may end up extending further.
In the final step we extend the patterns forward and find a square substring bounded by the ends of the resulting extensions, marked by vertical line segments in the bottom image.
This square substring may be different from $A$, but is always at least as long as $A$.}
\label{fig:lss-case1}
\end{figure}

Next, in a similar fashion, we find the largest positive integer $k\le \Delta-j$ such that $s(g, \dots, g+k] = s(h, \dd, h+k]$.
In this step we are maximally extending the copies of $P$ forwards while making sure they do not overlap with our previous extension.

Now, let $j'$ be the positive integer such that $A = s(g-j'\dd  g-j'+\ell]$ is the optimal solution whose first half and second half each contain the copies of $P$ we are considering. 
Then because $A$ is a square substring with shift $\Delta = h-g$, we have $s(g-j'\dd g] = s(h-j'\dd h]$, which implies that $j\ge j'$ by construction.
But since $A$ is square we also have $s(g\dd g + \Delta-j'] = s(h\dd h+\Delta-j']$.
Then the definition of $k$ together with the observation that $j\ge j'$ forces $k = \Delta - j$
(this is pictured in the bottom two images of \cref{fig:lss-case1}, where the left green substring and right blue substring are bordering each other).
Combining these observations together, we get that $s(g-j\dd h+k]$ is a square substring of size $2\Delta = \ell$.
Thus returning this substring produces the desired solution.

\paragraph*{Case 2: Multiple Copies.}
It remains to consider the case where $S$ contains multiple copies of $P$.
In this case, we use the quantum algorithm for Exact String Matching to find the rightmost and second rightmost copies of $P$ in $S$ in $\tilde{O}(\sqrt d)$ time.
Suppose that these copies start after positions $h$ and $h'$ respectively, so that 
\[P = s(h\dd h+2d/5] = s(h'\dd h'+2d/5]\]
with $h' < h$.
Then since $2|P| = 4d/5 > (3/5+\eps)d = |S|$, we know by \cref{lm:period-patterns} that $P$ has minimum period $p = h-h'$ and every appearance of $P$ in $S$ starts some multiple of $p$ away from  $h$. 
Moreover, all the copies of $P$ in $S$ overlap each other and together form one larger $p$-periodic substring in $s$.
Our next step will be to extend these periodic parts to maximal periodic substrings, which well help us locate a large square substring.

By Exact String Matching, we can find the leftmost copy $s(l\dd l+2d/5]$ of $P$ in $S$ in $\tilde{O}(\sqrt d)$, where the integer $l+1$ is the starting position of this copy.
By our earlier discussion, we know that the string $s(l\dd h+2d/5]$ is $p$-periodic.

We now extend the original pattern $P$ as well as the leftmost copy of $P$ in $S$ backwards while maintaining the property of being $p$-periodic. 

Formally, we find the largest nonnegative integer $j_1\le 2d/5$ such that $s(g-j_1\dd g+2d/5]$ is $p$-periodic and the largest nonnegative integer $j_2\le l+1-g-2d/5$ such that $s(l-j_2\dd h+2d/5]$ is $p$-periodic. 
Because we upper bound $j_1, j_2\le O(d)$, extending the strings in this way takes $\tilde{O}(\sqrt d)$ time via Grover search. 
We now split into two further subcases, depending on how far back the strings are extended.

\paragraph*{Case 2a: Single Periodic Substring.}

Suppose we get $j_2 = l+1-g-2d/5$.
This means that we were able to extend the leftmost copy of $P$ in $S$ so far back that it overlapped with our original pattern $P$ contained in the first half of $A$.
It follows that the substring $s(g\dd h+2d/5]$ is $p$-periodic. 
In particular, we deduce that the substring starting from the original pattern $P$ in the first half of $A$ to its $\Delta$-shifted copy in the second half of $A$ is contained in this $p$-periodic part.
Since $A$ is a square substring, it follows that its prefix which ends at position $g+2d/5$ of $s$ is also $p$-periodic. 
Thus, position $g-j_1$ of $s$ occurs before the first character of $A$.
This reasoning is depicted in the second image of \cref{fig:lss-case2a}.

%% Figure LSS Case 2a
\begin{figure}[t]
\centering
\def\svgwidth{\linewidth}
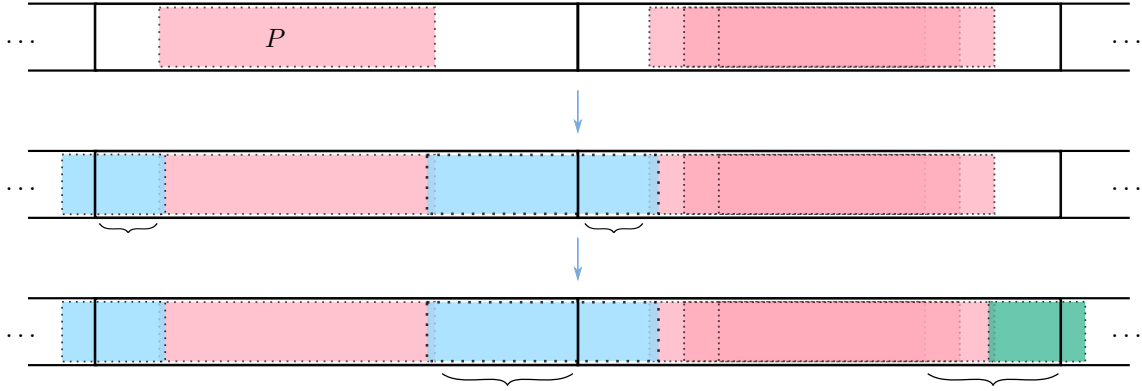
\caption{An example of how we extract a square substring when there are multiple copies of $P$ and the light blue substring spanning the copies of $P$ across both halves of $A$ is $p$-periodic.
In this case, because $A$ is square, when we extend the initial pattern $P$ backwards to a $p$-periodic string we cover a prefix of $A$. Similarly, when we extend copies of $P$ forward to a $p$-periodic string we cover a suffix of $A$. Curly braces indicate parts of the strings guaranteed to be identical because $A$ is square.}
\label{fig:lss-case2a}
\end{figure}

We now extend this entire $p$-periodic substring forward.
Find the largest nonnegative integer $k\le (1+\eps)d$ such that $s(g-j_1\dd g+k]$ is $p$-periodic. Since $j_1, k\le O(\log d)$ this takes $\tilde{O}(\sqrt d)$ time via Grover search.
As pictured in the bottom image of \cref{fig:lss-case2a}, since $A$ is a square string and the end of its first half is $p$-periodic, the end of its second half is $p$-periodic as well.
Thus position $g+k$ in $s$ occurs after the final character of $A$.

We now have a $p$-periodic string $s(g-j_1\dd g+k]$ which contains $A$, and thus has length at least $\ell$.
This means that $A$ has period $p$ as well.
We claim the shift $\Delta$ associated with $A$ is an integer multiple of $p$.

Indeed, by definition, $A$ is $\Delta$-periodic.
Then because $A$ has length $2\Delta \ge \Delta + p$, by \Cref{weak-period} we know that $A$ is $\gcd(p,\Delta)$-periodic as well.
Since $P$ is a substring of $A$, $P$ must have period $\gcd(p,\Delta)$ too.
But $p$ is the minimum period of $P$.
Hence $p = \gcd(p,\Delta)$, so $\Delta = pm$ for some positive integer $m$ as claimed. 

Thus $A$ has length $\ell = 2m\cdot p$.
Our $p$-periodic string $s(g-j_1\dd g+k]$ is guaranteed to be at least this long.
Thus, we can simply return the substring $s(g-j_1\dd g+2mp]$.
Because this substring is $p$-periodic and has length an even multiple of $p$, it is a square substring.
Because it has length equal to $A$, it is a longest square substring as desired.

\paragraph*{Case 2b: Disjoint Periodic Substrings.}

If we do not fall into {\bf Case 2a}, then we must have $j_2 < l+1-g-2d/5$,
so that the $p$-periodic substrings $s(g-j_1\dd g+2d/5]$ and $s(l-j_2\dd h+2d/5]$ do not overlap.
In this case, we construct two candidate solutions, and afterwards prove that one of them is guaranteed to be a longest square substring.

%% Figure LSS Case 2bi
\begin{figure}[t]
\centering
\def\svgwidth{\linewidth}
%% Creator: Inkscape 1.0.2 (e86c870879, 2021-01-15), www.inkscape.org
%% Accompanies image file 'qstring-case2bi.pdf' (pdf, eps, ps)
\begingroup%
  \makeatletter%
  \providecommand\color[2][]{%
    \errmessage{(Inkscape) Color is used for the text in Inkscape, but the package 'color.sty' is not loaded}%
    \renewcommand\color[2][]{}%
  }%
  \providecommand\transparent[1]{%
    \errmessage{(Inkscape) Transparency is used (non-zero) for the text in Inkscape, but the package 'transparent.sty' is not loaded}%
    \renewcommand\transparent[1]{}%
  }%
  \providecommand\rotatebox[2]{#2}%
  \newcommand*\fsize{\dimexpr\f@size pt\relax}%
  \newcommand*\lineheight[1]{\fontsize{\fsize}{#1\fsize}\selectfont}%
  \ifx\svgwidth\undefined%
    \setlength{\unitlength}{425.19685039bp}%
    \ifx\svgscale\undefined%
      \relax%
    \else%
      \setlength{\unitlength}{\unitlength * \real{\svgscale}}%
    \fi%
  \else%
    \setlength{\unitlength}{\svgwidth}%
  \fi%
  \global\let\svgwidth\undefined%
  \global\let\svgscale\undefined%
  \makeatother%
  \begin{picture}(1,0.13333333)%
    \lineheight{1}%
    \setlength\tabcolsep{0pt}%
    \put(0,0){\includegraphics[width=\unitlength,page=1]{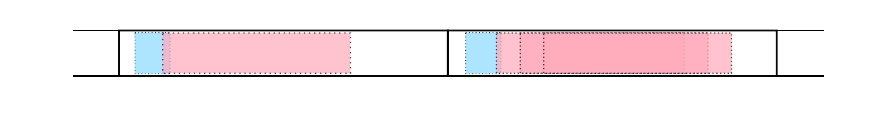}}%
    \put(0.0653867,0.06895509){\color[rgb]{0,0,0}\makebox(0,0)[lt]{\lineheight{1.25}\smash{\begin{tabular}[t]{l}$\dots$\end{tabular}}}}%
    \put(0.91558065,0.06895509){\color[rgb]{0,0,0}\makebox(0,0)[lt]{\lineheight{1.25}\smash{\begin{tabular}[t]{l}$\dots$\end{tabular}}}}%
    \put(0.2648867,0.06542842){\color[rgb]{0,0,0}\makebox(0,0)[lt]{\lineheight{1.25}\smash{\begin{tabular}[t]{l}$P$\end{tabular}}}}%
  \end{picture}%
\endgroup%

\caption{If extending $P$ backwards (pictured by the light blue substrings) to a $p$-periodic substring does not allow us to cover the prefix of $A$, then these $p$-periodic parts are identical because $A$ is square. 
Thus the distance $\Delta'$ between the beginnings of the light blue substrings equals the true shift $\Delta$.}
\label{fig:lss-case2bi}
\end{figure}
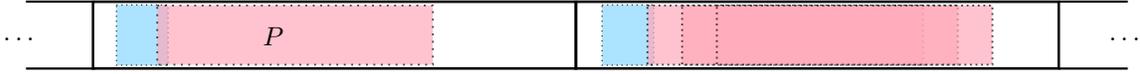

Define $\Delta' = (l-j_2) - (g-j_1)$.
Via Grover search in $\tilde{O}(\sqrt d)$ time, we find the largest nonnegative integers $b,b'\le \Delta'$ such that $s(g-j_1-b\dd g-j_1+1] = s(l-j_2-b\dd h-j_1+1]$ 
and $s(g-j_1\dd g-j_1+b'] = s(l-j_2\dd l-j_2+b']$.
We then set string $B = s(g-j_1-b\dd l-j_2+b']$ to be our first candidate solution.
Intuitively, this candidate corresponds to a guess that $\Delta = \Delta'$.

To construct the second candidate, we use a similar procedure, but first extend the strings forward.
Using Grover search, we find the largest positive integers $k_1\le l-j_2-g$ and $k_2\le (1+\eps)d$ such that $s(g\dd g+k_1]$ and $s(l\dd l+k_2]$ are each $p$-periodic, in $\tilde{O}(\sqrt d)$ time.
Set $\Delta'' = (l+k_2) - (g+k_1)$.
Then, as before, we find the largest nonnegative integers $c,c'\le \Delta''$ such that $s(g+k_1-c\dd g+k_1] = s(l+k_2-c\dd l+k_2]$ and $s(g+k_1\dd g+k_1+c'] = s(l+k_2\dd l+k_2+c']$.
The string $C = s(g+k_1-c\dd l+k_2+c']$
is then our second candidate.
Intuitively, this corresponds to a guess that $\Delta = \Delta''$.

We can check if $B$ and $C$ are square in $\tilde{O}(\sqrt d)$ time by Grover search.
If neither of them are square we report that we find no square substring.
Otherwise, we return the largest square substring among these two. 
It remains to prove that this procedure is correct.
There are two cases to consider, based off how large $j_1$ is relative to the position of $A$.

First, suppose that position $g-j_1+1$ in $s$ is a character in the first half of $A$.
Then, as depicted in \cref{fig:lss-case2bi}, since $A$ is square, $l-j_2+1$ must also be in the second half of $A$, and in fact be exactly $\Delta$ characters to the right of $g-j_1+1$ (because if this position was earlier, it would mean we could have picked $j_1$ larger and still had a $p$-periodic string).
Thus $\Delta = (l-j_2 + 1) - (g-j_1 + 1) = \Delta'$ is forced.
Then when we construct the string $B$ by searching backwards and forwards from positions $g-j_1+1$ and $l-j_2+1$ we will in fact find a square string of length $A$, and $B$ will our desired longest square substring.

Otherwise, position $g-j_1+1$ in $s$ is placed before every character of $A$.
Then as depicted in \cref{fig:lss-case2bii}, since $A$ is square, position $l-j_2$ must be in the first half of $A$.
Consequently, when we extend $P$ forward to position $g+k_1$, this position is also in the first half of $A$ (otherwise the $p$-periodic parts would overlap, and we would have been in {\bf Case 2a} instead). 
As in the previous case, using the fact that $A$ is a square again, we get that position $l+k_2$ must be exactly $\Delta$ characters to the right of $g+k_1$.
So $\Delta = (l+k_2) - (g + k_1) = \Delta''$ is forced.
Then when we construct the string $C$ by searching backwards and forwards from positions $g+k_1$ and $l+k_2$ we find a square string of length $A$, so $C$ will be our desired longest square substring.

%% Figure LSS Case 2bii
\begin{figure}[t]
\centering
\def\svgwidth{\linewidth}
%% Creator: Inkscape 1.0.2 (e86c870879, 2021-01-15), www.inkscape.org
%% Accompanies image file 'qstring-case2bii.pdf' (pdf, eps, ps)
\begingroup%
  \makeatletter%
  \providecommand\color[2][]{%
    \errmessage{(Inkscape) Color is used for the text in Inkscape, but the package 'color.sty' is not loaded}%
    \renewcommand\color[2][]{}%
  }%
  \providecommand\transparent[1]{%
    \errmessage{(Inkscape) Transparency is used (non-zero) for the text in Inkscape, but the package 'transparent.sty' is not loaded}%
    \renewcommand\transparent[1]{}%
  }%
  \providecommand\rotatebox[2]{#2}%
  \newcommand*\fsize{\dimexpr\f@size pt\relax}%
  \newcommand*\lineheight[1]{\fontsize{\fsize}{#1\fsize}\selectfont}%
  \ifx\svgwidth\undefined%
    \setlength{\unitlength}{425.19685039bp}%
    \ifx\svgscale\undefined%
      \relax%
    \else%
      \setlength{\unitlength}{\unitlength * \real{\svgscale}}%
    \fi%
  \else%
    \setlength{\unitlength}{\svgwidth}%
  \fi%
  \global\let\svgwidth\undefined%
  \global\let\svgscale\undefined%
  \makeatother%
  \begin{picture}(1,0.13333333)%
    \lineheight{1}%
    \setlength\tabcolsep{0pt}%
    \put(0,0){\includegraphics[width=\unitlength,page=1]{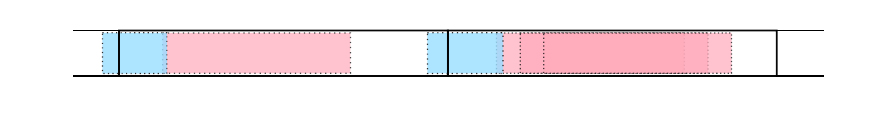}}%
    \put(0.0653867,0.06895509){\color[rgb]{0,0,0}\makebox(0,0)[lt]{\lineheight{1.25}\smash{\begin{tabular}[t]{l}$\dots$\end{tabular}}}}%
    \put(0.91558065,0.06895509){\color[rgb]{0,0,0}\makebox(0,0)[lt]{\lineheight{1.25}\smash{\begin{tabular}[t]{l}$\dots$\end{tabular}}}}%
    \put(0.2648867,0.06542842){\color[rgb]{0,0,0}\makebox(0,0)[lt]{\lineheight{1.25}\smash{\begin{tabular}[t]{l}$P$\end{tabular}}}}%
    \put(0,0){\includegraphics[width=\unitlength,page=2]{qstring-case2bii.pdf}}%
  \end{picture}%
\endgroup%

\caption{If in {\bf Case 2b}, extending $P$ backwards  (pictured by the light blue substring on the left) to a $p$-periodic substring covers a prefix of $A$, then when we extend $P$ forward in the same way (pictured by the left green substring) the result cannot cross into the second half of $A$ (if it did, we would have a single connected periodic substring and fall into {\bf Case 2a}). 
Then the $p$-periodic parts at the ends of each half must be identical because $A$ is square. Thus the distance $\Delta''$ between the ends of the green substrings equals the true shift $\Delta$.}
\label{fig:lss-case2bii}
\end{figure}

This handles all of the cases.
So far, we have a described an algorithm that, for any integer $i$, will find the longest square substring of $s$ with size in $[d,(1+\eps)d)$ with probability at least $\Omega(d/n)$ (recall this is the probability that $g$ is good), in time $\tilde{O}(\sqrt d)$.
By amplitude amplification and trying out the $O(\log n)$ choices of $i$ in decreasing order, we recover an algorithm for the Longest Square Substring problem which runs in \[\tilde{O}\left(\sqrt d \cdot \sqrt{n/d}\right) = \tilde{O}(\sqrt n)\] time, as desired.
\end{proof}

We show that our algorithm is optimal by giving a quantum query lower bound of $\Omega(\sqrt n)$ for finding the longest square substring.
This proof is essentially already present in \cite{legall}, where the authors give a lower bound for finding the longest \emph{palindromic} substring, but we sketch the argument here for completeness. 

\begin{prop}
    Any quantum algorithm that computes the longest square substring of a string of length $n$ requires $\Omega(\sqrt n)$ queries. 
\end{prop}
\begin{proof}
Let $S$ be the set of strings of length $2n$ over the alphabet $\set{0,1}$ which contain at most one occurrence of the character $1$.
In 
\cite{DBLP:journals/siamcomp/BennettBBV97}
the authors prove that deciding with whether a given string $s\in S$ is the string consisting of all $0$s requires $\Omega(\sqrt n)$ queries in the quantum setting. 

The longest square substring of the $0$s string of length $2n$ is just the entire string, and has length $2n$.
However, every other string in $S$ has an odd number of $1$s, and thus has longest square substring of size strictly less than $2n$.
So solving the Longest Square Substring problem lets us decide if a string from $S$ is the all $0$s string, which means that any quantum algorithm solving this problem requires $\Omega(\sqrt n)$ queries as well.
\end{proof}

\section{Open Problems}
\label{sec:open}
We conclude by mentioning several open questions related to our work.

\begin{itemize}
    \item Our $\tilde O(n^{2/3})$-time quantum algorithm for Longest Common Substring is near-optimal, as distinguishing between the case of $\mathsf{LCS}=0$ and $\mathsf{LCS} \ge 1$ requires $\Omega(n^{2/3})$ query complexity \cite[Section 7]{legall}. 
    However, for the problem of distinguishing between $\mathsf{LCS}\ge d$ and $\mathsf{LCS}< d$, this $\Omega(n^{2/3})$ query lower bound no longer applies.
    In fact, Le Gall and Seddighin \cite[Section 3.1.2]{legall} gave a quantum algorithm for this problem in $\tilde O(n/\sqrt{d})$ time, which is much faster than our $\tilde O(n^{2/3})$-time algorithm for large $d$, and is near-optimal for $d = \Theta(n)$. 
    Hence, an interesting open question is to find matching  upper bounds and lower bounds (in terms of both $n$ and $d$) for this decision problem when parameterized additionally by this threshold $d$. 
    For example, could there be an algorithm with $\tilde O(n^{2/3}/d^{1/6})$ query complexity (which would be optimal for both $d=1$ and $d=n$)? 
    \item In our time-efficient implementation of the LCS algorithm, we used a simple sampling technique to bypass certain restrictions on 2D range query data structures (\cref{sec:apply-ds}). Can this idea have further applications in designing time-efficient quantum walk algorithms? 
    
    As a simple example, we can use this idea to get an $\tilde O(n^{2/3})$-time comparison-based algorithm for the element distinctness problem with \emph{simpler implementation}.   At the beginning, uniformly sample $r$ items $x_1,\dots,x_r$ from the input array, and sort them so that $x_1\le \dots \le x_r$.
    Then, we create a hash table with $r+1$ buckets each having $O(\log n)$ capacity, where the hash function $h(x)$ is defined as the index $i$ such that $x_i\le x < x_{i+1}$, which can be found by binary search.
    Then, each insertion, deletion, and search operation can be performed in $O(\log n)$ time, provided that the buckets do not overflow. The error caused by overflows can be analyzed using Ambainis' proof of \cite[Lemma 6]{DBLP:journals/siamcomp/Ambainis07}.
    
    In comparison, Ambainis' implementation \cite{DBLP:journals/siamcomp/Ambainis07} additionally used a skip list, and Jeffery's  (non-comparison-based) implementation used a quantum radix tree \cite[Section 3.3.4]{jeffery2014frameworks}.
    
    \item Our $\tilde O(n^{2/3})$-time algorithm for LCS assumes that the input characters are integers in $[\poly(n)]$. This assumption was used for constructing string synchronizing sets sublinearly (\cref{sec:sync}).
    However, the previous $\tilde O(n^{5/6})$-time algorithm by Le Gall and Seddighin \cite{legall} can work with \emph{general ordered alphabet}, where the only allowed query is to compare two symbols $S[i],S[j]$ in the input strings (with three possible outcomes $S[i]>S[j],S[i]=S[j],$ or $S[i]<S[j]$).
    Is $\tilde O(n^{2/3})$ query complexity (or even time complexity) achievable in this more restricted setting? Alternatively, can  we show a better query lower bound? 
    \item Our algorithm  for the Minimal String Rotation problem (and other related problems in \cref{sec:lmsr}) has time complexity (and query complexity) $n^{1/2+o(1)}$.
    Can we reduce the $n^{o(1)}$ factor down to $\polylog(n)$?
    It seems like doing this within our current divide-and-conquer framework would be quite difficult, unless one can efficiently implement each level of the algorithm without any overhead.  Can we design a better algorithm that achieves the optimal $1/2$ exponent while only having a constant number of nested levels?
\end{itemize}

\section*{Acknowledgements}
We thank Virginia Vassilevska Williams, Ryan Williams, and Yinzhan Xu for several helpful discussions. 
We additionally thank Virginia Vassilevska Williams for several useful comments on the writeup of this paper.

	\bibliographystyle{alphaurl} 
	
	\bibliography{main}

\end{document}